\documentclass[journal]{IEEEtran}

\usepackage{cite}
\ifCLASSINFOpdf
\else
\fi
\usepackage{amsmath}
\usepackage{amssymb}

\usepackage{fontenc}
\usepackage{color}
\newcommand*\arc{{\fontfamily{pbk}\fontseries{db}\selectfont+}}

\newcommand\sbullet[1][.5]{\mathbin{\vcenter{\hbox{\scalebox{#1}{$\bullet$}}}}}

\newtheorem{theorem}{Theorem}
\newtheorem{lemma}{Lemma}

\usepackage{algorithm}
\usepackage{algorithmic}

\usepackage{subfigure}
\ifCLASSINFOpdf
   \usepackage[pdftex]{graphicx}
   \graphicspath{{../pdf/}{../jpeg/}}
   \DeclareGraphicsExtensions{.pdf,.jpeg,.png}
\else
   \usepackage[dvips]{graphicx}

   \graphicspath{{../eps/}}

   \DeclareGraphicsExtensions{.eps}
\fi
\ifCLASSOPTIONcompsoc
\else
  \usepackage[caption=false,font=footnotesize]{subfig}
\fi
\begin{document}

\title{Energy-Efficient Cooperative Secure Transmission in Multi-UAV Enabled Wireless Networks}

\author{Meng~Hua,~\IEEEmembership{Student Member,~IEEE,}
        Yi~Wang,
Qingqing~ Wu,~\IEEEmembership{Member,~IEEE,}
        Haibo~Dai,~\IEEEmembership{ Member,~IEEE,}
                Yongming~Huang,~\IEEEmembership{Senior Member,~IEEE}
        and~Luxi~Yang,~\IEEEmembership{Senior Member,~IEEE.}
\thanks{Copyright (c) 2015 IEEE. Personal use of this material is permitted. However, permission to use this material for any other purposes must be obtained from the IEEE by sending a request to pubs-permissions@ieee.org.}
\thanks{Manuscript received December  16, 2018; revised April   29, and June 12, 2019; accepted June 18, 2019. This work was supported by Scientific Research Foundation of Graduate School of Southeast University  under Grand  YBPY1859 and China Scholarship Council (CSC) Scholarship, National High Technology Project of China  under 2015AA01A703,  Cyrus Tang Foundation Endowed Young Scholar Program under SEU-CyrusTang-201801, Scientific and Technological Key Project of Henan Province under Grant 192102210246, Scientific Key Research Project of Henan Province for Colleges and Universities under Grand 19A510024, and China Postdoctoral Science Foundation under Grant 2018M633733 and 2019M651914,  National Natural Science Foundation of China under Grant  61801435, Grant 61671144, Grant 61801170, Grant 61801243. The associate editor coordinating the review of this paper and approving it for publication was Prof. Jingon Joung. (Corresponding author: Luxi Yang.)}
\thanks{M. Hua,  Y. Huang and L. Yang are with the School of Information Science and Engineering, Southeast University, Nanjing 210096, China (e-mail: \{mhua,  huangym, lxyang\}@seu.edu.cn).}
\thanks{Y. Wang is with School of Electronics and Communication Engineering, Zhengzhou University of Aeronautics, Zhengzhou 450046, China, and also with  the Physical Layer Security Laboratory, National Digital Switching System Engineering and Technological Research Center, Zhengzhou 450002, China(e-mail: yiwang@zua.edu.cn). }
\thanks{Q. Wu is  with the Department of Electrical and Computer Engineering, National University of Singapore, Singapore.(e-mail: elewuqq@nus.edu.sg). }
\thanks{H. Dai is with the School of Internet of Things, Nanjing University of Posts and Telecommunications, Nanjing 210003, China (e-mail: hbdai@njupt.edu.cn).}
}

\maketitle

\begin{abstract}
This paper investigates a multiple unmanned aerial vehicles (UAVs) enabled  cooperative secure transmission scheme in the presence  of multiple potential eavesdroppers. Specifically,   multiple source   UAVs (SUAVs) send confidential information to multiple legitimate  ground users and  in the meantime multiple jamming UAVs (JUAVs) cooperatively transmit interference signals to multiple eavesdroppers in order to improve the legitimate users' achievable secrecy rate.  By taking into account the limited energy budget of UAV, our goal is to maximize the system secrecy energy efficiency (SEE), namely the achievable secrecy rate per energy consumption unit, by jointly optimizing the UAV trajectory,  transmit power  and user scheduling under the constraints of UAV mobility as well as the maximum transmit power. The resulting optimization problem is  shown to be  a non-convex and mixed-integer fractional optimization problem, which is  challenging to solve. We decompose the original problem into three sub-problems, and  then an efficient iterative algorithm is proposed by leveraging the block coordinate descent and Dinkelbach method in combination with successive convex approximation  techniques. Simulation results show that the  proposed scheme outperforms the other benchmarks significantly in terms of the system SEE.
\end{abstract}


%
\IEEEpeerreviewmaketitle

\section{Introduction}
Historically, unmanned aerial vehicles (UAVs) are primarily used in military purposes, such as recognition mission, surveillance mission, territory detection and others \cite{filippone2006flight}.  The UAVs become more and more multi-functional and are promising for  civilian applications due  to their   flexibility  and cost-efficiency for wireless networks deployment \cite{hayat2016survey,Lyu2017Placement,Lyu2018uav,zeng2018trajectory}. The  UAV's flexible mobility can be exploited to design a trajectory that increases network throughput. In fact, there  has been considerable work on the implementation of UAV-enabled wireless communication systems such as UAV-aided relaying, UAV-aided ubiquitous coverage, UAV-aided information dissemination as well as data collection, etc \cite{zeng2016throughput,hua2018outage,alzenad20173,Mozaffari2016Efficient,bor2016efficient,hua2019joint,wu2018common, wu2018Joint,zhan2018energy,zeng2019energy}. Among them, the works in \cite{zeng2016throughput} and \cite{hua2018outage} considered that the UAV acted as the mobile relaying to deliver the source data to a destination node. In \cite{alzenad20173,Mozaffari2016Efficient,bor2016efficient,hua2019joint}, the UAV-aided wireless coverage problem was taken into account with the goal of either maximizing coverage region or maximizing user experience, i.e., quality of service (QoS). The works in \cite{wu2018common, wu2018Joint,zhan2018energy,zeng2019energy} studied the UAV-aided data dissemination and collection problems from the perspective of the common throughput maximization in either downlink or uplink transmission by optimizing UAV trajectory.

Although  the UAVs leveraged in wireless communication bring many benefits, e.g., higher throughput, better QoS, lower delay, etc.,  the UAV-aided wireless communication systems are not safe and are more vulnerable  to be wiretapped by the malicious  eavesdroppers due to the high probabilistic  LoS  of air-to-ground (A2G) channel \cite{zhang2019securing,cui2018robust,yang2018energy}. By far, the research of physical layer security (PLS) transmission in traditional cellular networks has been widely studied based on the design of maximizing the security rate \cite{zhu2017beamforming,Zhu2016Outage,zhu2018secure,zhang2015secure,zhang2016secure,cumanan2017secure,zhang2017secure}. For example, \cite{cumanan2017secure} studied the secrecy rate   maximization problem in the presence of multiple eavesdroppers. Simulation results showed that with the help of
 multiple jammers, the system secrecy rate was significantly improved. In \cite{zhang2017secure}, the authors further studied  a secure energy efficiency optimization problem in the cognitive radio  network under the constraints of the minimum harvested energy at energy receiver and the maximum interference leakage at the primary network.  It is worth noting that the transmit sources such as base station and WiFi access point are fixed. However, in the UAV-enabled PLS transmission systems, the UAV can freely adjust its heading  for executing the missions. If  the UAV flies closer to the eavesdroppers, the confidential information   is  more susceptible to be encrypted. Therefore, it imposes a new challenge to design  UAV-aided secure wireless networks. There have been a few works on the research of  the UAV-aided secure wireless networks \cite{zhang2017securing,wang2018joint, li2018UAV,cai2018dual,Lee2018UAV,zhong2018secure}.  Specifically, the work in \cite{zhang2017securing} considered a downlink transmission system where the  UAV sent the confidential message to a  legitimate ground user in the presence of a malicious ground eavesdropper, and a joint UAV trajectory and transmit power  was  proposed  to maximize the secrecy rate. In \cite{wang2018joint}, the authors considered  a UAV-aided mobile relying system to enhance the system's PLS by  jointly optimizing the UAV/source  transmit power and UAV trajectory. \cite{ li2018UAV} investigated a UAV-enabled mobile jamming scheme, where a UAV was leveraged to send  jamming signal to the potential eavesdropper for  maximizing  the average secrecy rate. Two UAVs  applied in  secure transmission systems were investigated in \cite{cai2018dual,Lee2018UAV},  where one UAV intended to send the confidential message and the other UAV  sent the jamming signal  by jointly optimizing the two UAVs trajectories as well as the transmit power to maximize secrecy rate. The work  \cite{zhong2018secure} further considered a worst-case secrecy rate maximization case by assuming that the two  UAVs only perfectly know the legitimate ground user while partially know the  eavesdropper's location.

Unfortunately,  the sustainability and performance of  UAV-enabled communication systems are fundamentally limited by the onboard energy of UAVs\cite{hua2018power,zeng2018energy}. Therefore,  the secrecy energy efficiency (SEE) of UAV-enabled communication systems, which defined as the ratio of the   secrecy rate to UAV energy  consumption measured by \text{bits/Joule}, is of paramount importance in practice.  In fact, there has considerable  work on the study of  the SEE in  traditional PLS systems \cite{sheng2018power,wang2018energy,xu2016secure,nghia2017mimo}. Our work is different from the previous studies, which considered the communication related energy. Note that  the UAV communication energy is several orders of  magnitude lower than the UAV propulsion energy consumption \cite{zeng2018energy}. Therefore, the design  of SEE problem in  UAV-enabled networks  still remains a new open issue. However, there only  one literature paid attention  to investigating it. In \cite{xiao2018secrecy}, a UAV is exploited to assist delivering the   confidential message from a ground source to a legitimate  node in the presence of a potential eavesdropper, in which the goal  was to maximize system SEE by jointly optimizing UAV trajectory, transmit  power,  and  communication scheduling. However,   work \cite{xiao2018secrecy} is only limited to relaying, in our paper, the UAVs used as aerial base stations are investigated.

In this paper, we consider a system  where multiple source UAVs (SUAVs) cooperatively transmit information to the legitimate ground users in the presence of multiple eavesdroppers. In addition, to improve the system  secrecy rate,  the multiple jamming  UAVs (JUAVs)  are leveraged to cooperatively send  the jamming signals to the eavesdroppers.  The motivation of using multi-UAV lies in the potential cooperative transmission ability to achieve better system performance, such as higher secrecy rate and lower access delay. To enhance the spectrum efficiency, we assume  that all the UAVs  share the same bandwidth to communicate with the ground legitimate  users in the downlink transmission. Thus, the UAVs transmit power  should be carefully designed to mitigate the co-channel  interference. More importantly, the UAV's  propulsion energy consumption is significantly influenced by its velocity  and acceleration. As a result,  our goal is to maximize the system SEE   by jointly optimizing the  UAV trajectory,  transmit power,  and user scheduling under the constraints of UAV mobility as well as maximum transmit power.  Note that the  work \cite{Lee2018UAV} investigated  the system with single  SUAV and single JUAV serving multiple legitimate ground users  while  does not considered the UAV's propulsion energy consumption. In fact, numerical results have shown that the benefits of proposed scheme compared with one single  SUAV and single JUAV case. To the best of our knowledge, there is no existing work to study the energy-efficient cooperative secure transmission in multi-UAV enabled wireless communication systems.  The main contributions are summarized as follows.
\begin{itemize}
\item We  propose an energy-efficient cooperative secure transmission scheme in the multi-UAV enabled wireless communication systems. We  take into account maximizing the system secrecy rate and minimizing the UAV propulsion energy consumption, and trade off them by formulating the SEE maximization problem  subject to the constraints of UAV mobility as well as maximum transmit power.

\item We develop a three-layer iterative algorithm to solve the non-convex and mixed integer fractional optimization problem by using  block coordinate descent, Dinkelbach method, and  successive convex approximation (SCA) techniques. Specifically, for any given UAV trajectory and transmit power, the user scheduling is optimally solved with  low computational  complexity. For any given  UAV trajectory and user scheduling, we propose an efficient  algorithm to optimize the UAV transmit power by  using the SCA techniques. For any given user scheduling and transmit power, the UAV trajectory is obtained by applying SCA techniques and Dinkelbach method.

\item The  multiple SUAVs are used to cooperatively transmit information   to the legitimate users,  as expected, the achievable rate of UAV systems  can be  improved   by optimizing user scheduling and UAVs transmit power. Moreover, for enhancing  the performance of UAV-enabled secure transmission  systems, multiple JUAVs are leveraged  to transmit the jamming signal to the eavesdroppers.

\item Numerical results demonstrate the following conclusions.  First, the proposed  JUAVs-aided  scheme  achieves significantly higher secrecy rate compared with no JUAVs-aided scheme. Second, the system  SEE  does not   monotonically increase or decrease  with period time $T$, in contrast, the UAV system SEE   firstly increases with period time $T$ and then decreases with period time $T$. Third, the proposed scheme  outperforms  the other  benchmarks in terms of SEE.
\end{itemize}

The rest of this paper is organized as follows. In Section II, we introduce the multi-UAV enabled cooperative secure communication systems and formulate the SEE  problem. Section III  proposes an efficient iterative algorithm to solve the formulated problem by using  the block coordinate descent and Dinkelbach method, as well as SCA  techniques.  In Section IV,  numerical results are presented to illustrate superiority of our scheme. Finally, we  conclude the paper in Section V.
\begin{figure}[!t]
\centerline{\includegraphics[width=2.8in]{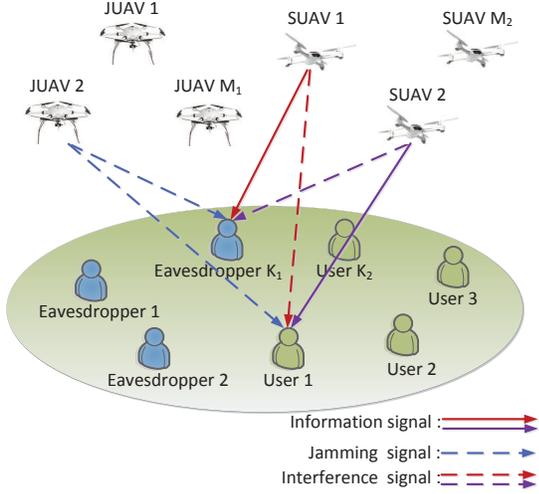}}
\caption{An illustration of a multi-UAV enabled cooperative secure system. } \label{fig1}
\end{figure}
\section{system model and problem formulation }
Consider a multiple UAVs-aided downlink secure transmission systems, which consist of a set ${\cal K}_2$ of $K_2$ legitimate users and a set  ${\cal K}_1$ of  $K_1$ eavesdroppers as shown in Fig.~\ref{fig1}. Denote a set of  $M_1$ JUAVs as ${\cal M}_1$ and a set of  $M_2$ SUAVs as ${\cal M}_2$. We assume that all the UAVs fly at a   fixed altitude $H$, which can be considered as the minimum altitude to avoid collision by the  infrastructure  obstacles. For simplicity, the  collision among UAVs is not considered here since  it can be readily extended by adding the minimum security distance constraints as in \cite{wu2018Joint}. In addition, we assume that SUAVs and JUAVs perfectly know  the legitimate user and eavesdropper locations by  proper information exchange.  This paper  can be easily extended to the case with imperfect location of the eavesdropper. Since our main focus is to present the proposed scheme based on UAV's SEE to the reader in an easy-to-access manner, we would like to restrict the current contribution to the location perfectly known case.

For ease of exposition,  a Cartesian coordinate system is established with all the dimensions measured in meters. The horizontal coordinates of the legitimate users $k_2 \in {\cal K}_2$ and the eavesdroppers $k_1 \in {\cal K}_1$ are respectively  denoted by  ${\bf w}_{k_2}$ and ${\bf w}_{k_1}$. The continuous time $T$ is equally divided into $N$ time slots with duration $\delta$. As such,   the horizontal coordinate of the $i$-th UAV's trajectory, velocity  and acceleration over horizon time $T$ can be approximately denoted by $N$\text{-}length sequences as $\{{\bf q}_i[n]\}$, $\{{\bf v}_i[n]\}$ and $\{{\bf a}_i[n]\}$, $i \in {\cal M}_1 \cup {\cal M}_2, n\in {\cal N}=\{1,2,...,N\}$.

Different from terrestrial communications, the UAV-to-ground channel is more likely to be dominated by LoS link, especially for rural or sub-urban environment.  Recent field experiments by Qualcomm have verified that the UAV-to-ground channel is indeed dominated by the LoS link for UAVs flying above a certain altitude \cite{qualcomm2017unmanned}. In addition, the LoS of A2G channel model is also one of the considered channel models in the recent 3GPP specification \cite{3GPP}. Therefore, the air-to-ground channel power gain from UAV $i$, $i \in {\cal M}_1 \cup {\cal M}_2$, to any ground user $k$, $k \in {\cal K}_1 \cup {\cal K}_2$, at time slot $n$ can be modeled as \cite{zeng2016throughput},\cite{wu2018Joint},\cite{matolak2017air},
\begin{align}
{h_{k,i}}\left[ n \right] = \frac{{{\beta _0}}}{{d_{k,i}^2\left[ n \right]}} = \frac{{{\beta _0}}}{{{{\left\| {{\bf{q}}_i\left[ n \right] - {{\bf{w}}}}_k \right\|}^2} + {H^2}}},
\end{align}
where  ${d_{k,i}\left[ n \right]}$ denotes the distance between UAV $i$ and ground user $k$ within time slot $n$, $\beta_0$ represents the reference channel gain at $d=1 \rm{m}$.
\subsection{Secrecy rate }
Define a binary variable $x_{k_2,m_2}$, which stands for that the legitimate  user $k_2$, $k \in {\cal K}_2$, is served by SUAV $m_2$, $m_2\in{\cal M}_2$, within time slot $n$ if  $x_{k_2,m_2}=1$, otherwise, $x_{k_2,m_2}=0$. Moreover, we assume that within each time slot $n$, one legitimate user  can be at most served by one SUAV, and each SUAV can  serve at most one legitimate user. Thus, we have
\begin{align}
&\sum\limits_{{k_2} = 1}^{{K_2}} {{x_{{k_2},{m_2}}}\left[ n \right]}  \le 1, {m_2}\in {\cal M}_2, n \in  {\cal N},  \label{scheduling1}\\
&\sum\limits_{{m_2} = 1}^{{M_2}} {{x_{{k_2},{m_2}}}\left[ n \right]}  \le 1, {k_2}\in {\cal K}_2,,n \in  {\cal N}, \label{scheduling2}\\
&{x_{{k_2},{m_2}}}\left[ n \right] \in \left\{ {0,1} \right\},\forall {k_2},{m_2},n \in  {\cal N} \label{scheduling3}.
\end{align}
If the legitimate user $k_2 \in {\cal K}_2$ is served by UAV $m_2 \in {\cal M}_2$ at  time slot $n \in {\cal N}$, the achievable rate of user $k_2$ in $\rm{bps}$ can be expressed as
\begin{align}
{R_{{k_2}}}[n] = \sum\limits_{{m_2} = 1}^{{M_2}} {{x_{{k_2},{m_2}}}\left[ n \right]{R_{{k_2},{m_2}}}\left[ n \right]},
\end{align}
where ${R_{{k_2},{m_2}}}\left[ n \right] = B{\log _2}\left( {1 + \frac{{{p_{{m_2}}}\left[ n \right]{h_{{k_2},{m_2}}}\left[ n \right]}}{{\sum\limits_{i \in {{\cal M}_2}\backslash \left\{ {{m_2}} \right\} \cup {{\cal M}_1}} {{p_i}\left[ n \right]{h_{{k_2},i}}\left[ n \right]}  + {\sigma ^2}}}} \right)$, $p_{m_2}[n]$ denotes the transmit power of SUAV $m_2$ at time slot $n$, and $\sigma ^2$ is the additive Gaussian white noise power.

Similarly, the achievable rate of eavesdropper $k_1$ wiretapping  the   user $k_2$ at time slot $n$ is given by
\begin{align}
{R_{{k_1} \to {k_2}}}[n] = \sum\limits_{{m_2} = 1}^{{M_2}} {{x_{{k_2},{m_2}}}\left[ n \right]{R_{{k_1},{m_2}}}\left[ n \right]}, \label{secrecyratetimeslot}
\end{align}
where ${R_{{k_1},{m_2}}}\left[ n \right]= B{\log _2}\left( {1 + \frac{{{p_{{m_2}}}\left[ n \right]{h_{{k_1},{m_2}}}\left[ n \right]}}{{\sum\limits_{i \in {{\cal M}_2}\backslash \left\{ {{m_2}} \right\} \cup {{\cal M}_1}} {{p_i}\left[ n \right]{h_{{k_1},i}}\left[ n \right]}  + {\sigma ^2}}}} \right)$.

Therefore, the worst-case secrecy rate of user $k_2 \in{\cal K}_2$ in the presence of $K_1$ eavesdroppers over horizon time $T$ is given by \cite{zhu2017beamforming}, \cite{Zhu2016Outage}
\begin{align}
R_{{k_2}}^{\sec } = {\sum\limits_{n = 1}^N {\left[ {{R_{{k_2}}}\left[ n \right] - \mathop {\max }\limits_{{k_1} \in {{\cal K}_1}} {R_{{k_1} \to {k_2}}}[n]} \right]} ^ + },
\end{align}
where $\left[b\right]^{+}={\rm max}\{b,0\}$.
\subsection{UAV propulsion energy consumption }
Herein, a fixed-wing UAV is employed since its flight endurance is typically much longer than that of the rotary-wing UAV. The total energy consumption of fixed-wing UAV consists of two parts: UAV communication energy and UAV propulsion energy. However, \cite{zeng2018energy} shows that the UAV communication energy consumption is several orders of magnitudes lower than the UAV propulsion energy consumption. Thus, the UAV communication energy consumption can be ignored compared with UAV propulsion energy consumption. Based on \cite{zeng2018energy}, the total propulsion energy consumption of UAV $i$ over $T$ is given by
\begin{align}
{E_i} = \sum\limits_{n = 1}^N {\left( {{c_1}{{\left\| {{{\bf{v}}_i}\left[ n \right]} \right\|}^3} + \frac{{{c_2}}}{{\left\| {{{\bf{v}}_i}\left[ n \right]} \right\|}}\left( {1 + \frac{{{{\left\| {{{\bf{a}}_i}\left[ n \right]} \right\|}^2}}}{{{g^2}}}} \right)} \right)},
\end{align}
where $g$ is the gravitational acceleration with nominal value $9.8{\rm m/s^2}$, $c_1$ and $c_2$ are the constant parameters related to the UAV wing area, air density and UAV's weight \cite{zeng2018energy}.
\subsection{Problem formulation }
Let us define $X = \left\{ {{x_{{k_2},{m_2}}}\left[ n \right],\forall {k_2},{m_2},n} \right\}$, $P = \left\{ {{p_{{i}}}\left[ n \right],\forall i,n} \right\}$, $Q = \left\{ {{{\bf{q}}_i}[n],{{\bf{v}}_i}[n],{{\bf{a}}_i}[n],\forall i,n} \right\}$ and ${\cal N}_1=\{0,1,...,N\}$. Mathematically, the problem can be formulated as follows
\begin{align}
&({\rm P})\mathop {\max }\limits_{X,P,Q} \frac{{\sum\limits_{{k_2} = 1}^{{K_2}} {{\sum\limits_{n = 1}^N {\left[ {{R_{{k_2}}}\left[ n \right] - \mathop {\max }\limits_{{k_1} \in {{\cal K}_1}} {R_{{k_1} \to {k_2}}}[n]} \right]} ^ + }} }}{{{E_{{\rm{total }}}}}}\notag\\
&{\rm s.t.}\quad \eqref{scheduling1},\eqref{scheduling2},\eqref{scheduling3},\\
& \qquad{{\bf{q}}_i}\left[ {n + 1} \right] = {{\bf{q}}_i}\left[ {n } \right] + {{\bf{v}}_i}\left[ n \right]\delta  + \frac{1}{2}{{\bf{a}}_i}\left[ n \right]{\delta ^2},\notag\\
&\qquad\qquad\qquad\qquad\qquad\qquad n\in  {\cal N}_1,i \in {{\cal M}_1} \cup {{\cal M}_2},\label{trajectory1}\\
&\qquad{{\bf{v}}_i}\left[ n+1 \right] = {{\bf{v}}_i}\left[ n \right] + {{\bf{a}}_i}\left[ n \right]\delta , n \in  {\cal N}_1,i \in {{\cal M}_1} \cup {{\cal M}_2},\label{trajectory2}\\
&\qquad{{\bf{q}}_i}\left[ 0 \right] = {{\bf{q}}_i}\left[ {N+1 } \right],i \in {{\cal M}_1} \cup {{\cal M}_2},\label{trajectory3}\\
&\qquad{{\bf{v}}_i}\left[ 0 \right] = {{\bf{v}}_i}\left[ {N+1 } \right],i \in {{\cal M}_1} \cup {{\cal M}_2},\label{trajectory4}\\
&\qquad\left\| {{{\bf{v}}_i}\left[ n \right]} \right\| \le {v_{\max }}, n \in  {\cal N}_1, i \in {{\cal M}_1} \cup {{\cal M}_2},\label{trajectory5}\\
&\qquad\left\| {{{\bf{a}}_i}\left[ n \right]} \right\| \le {a_{\max }},n \in  {\cal N}_1, i \in {{\cal M}_1} \cup {{\cal M}_2},\label{trajectory6}\\
&\qquad0 \le {p_i}\left[ n \right] \le {P_{\max }},i \in {{\cal M}_1} \cup {{\cal M}_2},\label{power}
\end{align}
where ${E_{\rm total}} = \sum\limits_{i \in {{\cal M}_1} \cup {{\cal M}_2} } {{E_i}}$;  \eqref{trajectory1}-\eqref{trajectory6} denote the UAV trajectory constraints; ${\bf q}_i[0]$ and  ${\bf q}_i[N+1]$ denote the initial location and final location of UAV $i$ ;  ${\bf v}_i[0]$ and ${\bf v}_i[N+1]$ denote the initial velocity  and final velocity of UAV $i$; \eqref{trajectory5}, \eqref{trajectory6},  and \eqref{power} represent the  feasible and boundary constraints of the optimization variables.
\section{energy efficient algorithm design for multi-UAV secure transmission }
The problem $(\rm P)$ is a non-convex and mixed integer fractional optimization problem, which is challenging to solve due to the  following reasons. First, the binary variables are involved in objective function and  constraints \eqref{scheduling1}-\eqref{scheduling3}. Second, both numerator and  denominator in objective function are non-convex, which lead to a non-convex fractional objective function. With that in mind, we decompose the original problem into three sub-problems, namely user scheduling optimization, UAV transmit power  optimization, and UAV trajectory optimization. Then, an efficient iterative algorithm is proposed via alternately optimizing these three sub-problems.
\subsection{User scheduling optimization}
In this subsection, we consider the first sub-problem for optimizing the  user scheduling $X$ with given UAV transmit power $P$ and UAV trajectory $Q$. Then, the problem $(\rm P)$ can be simplified as
\begin{align}
&({\rm{P1}})\mathop {\max }\limits_{X,{\Gamma _{{k_2}}}\left[ n \right]} \sum\limits_{{k_2} = 1}^{{K_2}} {\sum\limits_{n = 1}^N {{\Gamma _{{k_2}}}\left[ n \right]} } \notag\\
&{\rm s.t.}~\eqref{scheduling1}\text{-}\eqref{scheduling3},\notag\\
&\qquad{R_{{k_2}}}\left[ n \right] - \mathop {\max }\limits_{{k_1} \in {{\cal K}_1}} {R_{{k_1} \to {k_2}}}[n] \ge {\Gamma _{{k_2}}}\left[ n \right],n \in  {\cal N},k_2,\label{p1const1}
\end{align}
where $\Gamma_{k_2}[n]$ is the slack variable. It is observed that the  problem $(\rm P1)$ is an integer optimization problem, which in general has no  efficient algorithm to solve it with low computational complexity. A traditional method is to first relax the binary variables into continuous variables, and then reconstruct  the binary solution by using the obtained continuous solution \cite{wu2018Joint}.  However, it cannot be guaranteed that the  reconstructed binary solution  is the  optimal solution to  problem $(\rm P1)$. In addition, mapping the continuous solution to the optimal binary soltuion is a very challenging task, which has higher computational complexity. In contrast, we  can obtain the optimal scheduling to problem $(\rm P1)$ based on the Algorithm 1.
%

\begin{algorithm}
\caption{Proposed method to solve $({\rm P}1)$}
\begin{algorithmic}[1]
\STATE  \textbf{for} $k_2=1$ to $K_2$ \textbf{do} \\
\STATE  \quad   \textbf{for} $m_2=1$ to $M_2$ \textbf{do}\\
\quad \quad \textbf{if} $R_{{k_2},{m_2}}^{\sec }\left[ n \right]<0$\\
\quad \quad \quad  Set $x_{{k_2},{m_2}}\left[ n \right]=0$,  $R_{{k_2},{m_2}}^{\sec }\left[ n \right]=0$.\\
\quad \quad \textbf{end if}\\
\STATE \quad  \textbf{end for}\\
\STATE  \textbf{end for}\\

\STATE  \textbf{for} $k_2=1$ to $K_2$ \textbf{do} \\
\quad  Step 1: compute $\left\{ {{k_2},m_2^ * } \right\} = \mathop {\arg \max }\limits_{{m_2}} R_{{k_2},{m_2}}^{\sec }\left[ n \right]$, and \\
\quad set $x_{{k_2},{m_2}}\left[ n \right] = 0,{m_2} \in {{\cal M}_2}\backslash \left\{ {m_2^ * } \right\}$.
\STATE  \textbf{end for}\\
\STATE \textbf{if} $\{R_{{k_2},{m_2^*}}^{\sec }\left[ n \right],k_2\in {\cal K}_2\}==0$ \\
\quad    Step 2: set $x_{{k_2},{m_2^*}}\left[ n \right]=0$, and remove the $R_{{k_2},{m_2^*}}^{\sec }\left[ n \right]$ \\
\quad  from the set $\{R_{{k_2},{m_2^*}}^{\sec }\left[ n \right],k_2\in {\cal K}_2\}$.\\
\STATE \textbf{end if}
\STATE  Define  the new set $\{R_{{k_2},{m_2^*}}^{\sec }\left[ n \right],k_2\in {\cal K}_2^{new},m_2^*\in {\cal M}_2^{new}\}$.\\
\STATE \textbf{if}  there are  more than  two   same  index ${\bar m}_2^y$ in the indexes $\{{\bar m}_2^y\in {\cal M}_2^{new}\}$, \\
\quad Step 3: compute $\left\{ {{\bar k}_2^y,{\bar m}_2^y } \right\} = \mathop {\arg \max }\limits_{{k_2} \in  {\cal K}_{{\bar m}_2^y}} R_{{k_2},{{\bar m}_2}^y}^{\sec }\left[ n \right]$ \\
\quad (${\cal K}_{{\bar m}_2^y}$ is the user set containing ${\bar m}_2^y$).\\
\quad Step 4:  set ${x_{{\bar k}_2^y,{\bar m}_2^y}} = 1$, ${x_{k_2,{\bar m}_2^y}} = 0$ $({k_2} \in {{\cal K}_{\bar m_2^y}}\backslash \left\{ {\bar k_2^y} \right\})$, \\
\quad   and also remove $R_{k_2,{\bar m}_2^y}^{\sec }\left[ n \right]$  $({k_2}\in {{\cal K}_{\bar m_2^y}}\backslash \left\{ {\bar k_2^y} \right\})$ from  \\
\quad  the  set $\{R_{{k_2},{m_2^*}}^{\sec }\left[ n \right],k_2\in {\cal K}_2^{new},m_2^*\in {\cal M}_2^{new}\}$.
\STATE \textbf{end if}
\STATE Define the new set  $\{R_{{{\tilde k}_2},{{\tilde m}_2}}^{\sec}\left[ n \right],{\tilde k}_2\in {\cal K}_2^{renew},{\tilde m}_2\in {\cal M}_2^{renew}\}$, and set $x_{{\tilde k}_2,{\tilde m}_2}=1$ for ${\tilde k}_2\in {\cal K}_2^{renew},{\tilde m}_2\in {\cal M}_2^{renew}\}$.\\
\end{algorithmic}
\end{algorithm}
In Algorithm 1, $R_{{k_2},{m_2}}^{\sec }\left[ n \right] = {R_{{k_2,m_2}}}\left[ n \right] - \mathop {\max }\limits_{{k_1} \in {{\cal K}_1}} {R_{{k_1,m_2}}}[n]$, $\forall n, k_2 \in {\cal K}_2, m_2 \in {\cal M}_2$. Since the problem $({\rm P}1)$ can be split into  $N$ parallel sub-problems, thus we can solve it separately. For any given time slot $n$ in the Algorithm 1, step 1 stands for  that each legitimate user associated with  the SUAV with its  local benefits maximization. In fact,  step 1 obeys the constraint (3).  Step 2 makes sure that the secrecy rate is larger than zero when the user is scheduled. Step 3 and Step 4 complies with  the constraint (2).

For the stage 1\text{-}4, the computational complexity for this loop is $K_2M_2$. For the stage 5\text{-}6, in each loop, the complexity for selecting the optimal $m_2$ that maximizing $R_{{k_2},{m_2}}^{\sec }\left[ n \right]$ is $M_2-1$, and  thus the computational complexity for  the stage 5\text{-}6 is $K_2(M_2-1)$. For the stage 7\text{-}8, the computational complexity for the determine statement is $K_2$. For the  stage 10\text{-}11, the computational complexity for checking the same ${\bar m}_2^y$ is $C_{{\bar m}_2^y}N_{{\bar m}_2^y}$, where  $C_{{\bar m}_2^y}$ is the number of clusters with each cluster containing more than two same ${\bar m}_2^y$ $({\bar m}_2^y\in {\cal M}_2^{new})$, and $N_{{\bar m}_2^y}$  is the cardinality of the corresponding  cluster.  Note that $C_{{\bar m}_2^y}N_{{\bar m}_2^y} \le M_2$. The total computational complexity for solving $({\rm P}1)$ by using Algorithm 1 is ${\cal O}\left( {N\left( {2{K_2}{M_2}  + {C_{\bar m_2^y}}{N_{\bar m_2^y}}} \right)} \right)$.

However, for the  exhaustive search method, at any time slot $n$, there are ${{\mathbb P}\left( {{M_2},{K_2}} \right)}$  scheduling choices ($\mathbb P$ is a permutation operator).  We  compute the objective function results of problem $(\rm P1)$ with these choices (the complexity is ${{\mathbb P}\left( {{M_2},{K_2}} \right)}$), and then choose the optimal one that maximize problem $(\rm P1)$ (the complexity is ${{\mathbb P}\left( {{M_2},{K_2}} \right)}-1$). Thus, the total computational complexity for solving $({\rm P}1)$ by using exhaustive search  method is  ${\cal O}\left( {N \times \left( {2{\mathbb P}\left( {{M_2},{K_2}} \right) - 1} \right)} \right)\left( {{K_2} \ge {M_2}} \right)$. Obviously,  the computational complexity of our proposed method is much lower than the exhaustive search method, especially when  $N$,  $K_2$, and $M_2$ are large.


\subsection{UAV power optimization}
In this subsection, we consider the second sub-problem $(\rm P2)$ for optimizing UAV transmit power $P$ with given UAV trajectory $Q$ and user scheduling $X$, which can be written as
\begin{align}
&({\rm P}2)\mathop {\max }\limits_{{P,}{\tau _{{k_2}}}\left[ n \right],{\gamma _{ {k_2}}}\left[ n \right]} \sum\limits_{{k_2} = 1}^{{K_2}} {\sum\limits_{n = 1}^N {{\tau _{{k_2}}}\left[ n \right]} }\notag\\
&{\rm{s}}{\rm{.t}}{\rm{.}}{\kern 1pt} {\kern 1pt} {\kern 1pt} {\kern 1pt} {\kern 1pt} {\kern 1pt} {R_{{k_2}}}\left[ n \right] - {\gamma _{ {k_2}}}\left[ n \right] \ge {\tau _{{k_2}}}\left[ n \right],n \in  {\cal N},k_2, \label{p2const1}\\
&\qquad{\gamma _{ {k_2}}}\left[ n \right] \ge {R_{k_1 \to {k_2}}},n \in  {\cal N},{k_1},k_2,\label{p2const2}\\
&\qquad0 \le {p_i} \le {P_{\max }},i \in {{\cal M}_1} \cup {{\cal M}_2},
\end{align}
where $\tau_{k_2}$ and $\gamma_{k_2}$ are the auxiliary  variables.

Problem $(\rm P2)$ is a non-convex optimization problem owing to the non-convex constraints \eqref{p2const1}  and \eqref{p2const2}. To proceed, we resort to leveraging the SCA technique to obtain an efficient approximation solution to problem $(\rm P2)$. First, to tackle the non-convex  term ${R}_{k_2}[n]$  with respect to (w.r.t.) $p_i[n]$ in constraint \eqref{p2const1}, we apply the SCA technique to transform ${ R}_{k_2}[n]$ into a convex form. Then, we can rewrite $R_{k_2,m_2}$ as follows
\begin{align}
{R_{{k_2},{m_2}}}\left[ n \right] = {{\bar R}_{{k_2},{m_2}}}\left[ n \right] - {{\tilde R}_{{k_2},{m_2}}}\left[ n \right], {k_2} \in {{\cal K}_2}, \label{p2_1}
\end{align}
where ${{\bar R}_{{k_2},{m_2}}}\left[ n \right] = B{\log _2}\left( {\sum\limits_{i \in {{\cal M}_1} \cup {{\cal M}_2}} {{p_i}\left[ n \right]{h_{{k_2},i}}\left[ n \right]}  + {\sigma ^2}} \right)$ and ${{\tilde R}_{{k_2},{m_2}}}\left[ n \right] = B{\log _2}\left( {\sum\limits_{i \in {{\cal M}_2}\backslash \left\{ {{m_2}} \right\} \cup {{\cal M}_1}} {{p_i}\left[ n \right]{h_{{k_2},i}}\left[ n \right]}  + {\sigma ^2}} \right)$. It can be seen from \eqref{p2_1} that  $R_{k_2,m_2}$ is a difference of two concave functions. To proceed, by defining the set $\{p_i^r[n]\}$ as the given local point  at the $r$\text{-th} iteration, we have  the following result.
\newcounter{mytempeqncnt1}
\begin{figure*}
\normalsize
\setcounter{mytempeqncnt1}{\value{equation}}
\begin{align}
{{\tilde R}_{{k_2},{m_2}}}\left[ n \right] \le & B{\log _2}\left( {\sum\limits_{i \in {{\cal M}_2}\backslash \left\{ {{m_2}} \right\} \cup {{\cal M}_1}} {p_i^r\left[ n \right]{h_{{k_2},i}}\left[ n \right]}  + {\sigma ^2}} \right)\notag\\
&+B\sum\limits_{i \in {{\cal M}_2}\backslash \left\{ {{m_2}} \right\} \cup {{\cal M}_1}} {\frac{{{h_{{k_2},i}}\left[ n \right]{{\log }_2}\left( e \right)}}{{\sum\limits_{j \in {{\cal M}_2}\backslash \left\{ {{m_2}} \right\} \cup {{\cal M}_1}} {p_j^r\left[ n \right]{h_{{k_2},j}}\left[ n \right]}  + {\sigma ^2}}}} \left( {{p_i}\left[ n \right] - p_i^r\left[ n \right]} \right)\overset{a} = \tilde R_{{k_2},{m_2}}^{up}\left[ n \right].\label{tildeup}
\end{align}
\hrulefill 
\vspace*{4pt} 
\end{figure*}

\begin{lemma}
For any given power $\{p_i^r[n]\}$, the  inequality is hold in \eqref{tildeup}  (at the top of next page). \label{lemma2}
\end{lemma}
\begin{IEEEproof}
Please refer to Appendix A.
\end{IEEEproof}

From Lemma \ref{lemma2}, we can see that the $\tilde R_{k_2,m_2}^{up}[n]$ is linear w.r.t. the power $p_i[n]$, which indicates that $\tilde R_{k_2,m_2}[n]$ can be replaced by its  upper bound   $\tilde R_{k_2,m_2}^{up}[n]$. Thus, it can be readily verified that the constraint \eqref{p2const1} is now convex.

To address the non-convex constraint \eqref{p2const2}, the SCA technique is still applied. Similarly, for $k_1 \in {\cal K}_1$, we can reexpress $R_{k_1,m_2}[n]$ as
\begin{align}
{R_{{k_1},{m_2}}}\left[ n \right] = {{\bar R}_{{k_1},{m_2}}}\left[ n \right] - {{\tilde R}_{{k_1},{m_2}}}\left[ n \right], {k_1} \in {{\cal K}_1} \label{p2_2}
\end{align}
where ${{\bar R}_{{k_1},{m_2}}}\left[ n \right] = B{\log _2}\left( {\sum\limits_{i \in {{\cal M}_1} \cup {{\cal M}_2}} {{p_i}\left[ n \right]{h_{{k_1},i}}\left[ n \right]}  + {\sigma ^2}} \right)$ and ${{\tilde R}_{{k_1},{m_2}}}\left[ n \right] = B{\log _2}\left( {\sum\limits_{i \in {{\cal M}_2}\backslash \left\{ {{m_2}} \right\} \cup {{\cal M}_1}} {{p_i}\left[ n \right]{h_{{k_1},i}}\left[ n \right]}  + {\sigma ^2}} \right)$.  It can be seen that    $R_{k_1,m_2}$ is also a difference of two concave function, which brings the following lemma.
\begin{lemma}
For any given power $\{p_i^r[n]\}$, the inequality is hold in \eqref{barup}  (at the top of next page). \label{lemma3}
\end{lemma}
\begin{IEEEproof}
The proof is similar to Lemma \ref{lemma2} and omitted here for simplicity.
\end{IEEEproof}
\newcounter{mytempeqncnt2}
\begin{figure*}
\normalsize
\setcounter{mytempeqncnt2}{\value{equation}}
\begin{align}
{{\bar R}_{{k_1},{m_2}}}\left[ n \right] \le & B{\log _2}\left( {\sum\limits_{i \in {{\cal M}_1} \cup {{\cal M}_2}} {{p_i^r}\left[ n \right]{h_{{k_1},i}}\left[ n \right]}  + {\sigma ^2}} \right)\notag\\
&+ B\sum\limits_{i \in {{\cal M}_1} \cup {{\cal M}_2}} {\frac{{{h_{{k_1},i}}\left[ n \right]{{\log }_2}\left( e \right)}}{{\sum\limits_{j \in {{\cal M}_1} \cup {{\cal M}_1}} {p_j^r\left[ n \right]{h_{{k_1},j}}\left[ n \right]}  + {\sigma ^2}}}} \left( {{p_i}\left[ n \right] - p_i^r\left[ n \right]} \right) \overset{b} = \bar R_{{k_1},{m_2}}^{up}\left[ n \right].\label{barup}
\end{align}
\hrulefill 
\vspace*{4pt} 
\end{figure*}

Therefore, with the given local point $\{p_i^r[n]\}$ and upper bound results ${\tilde R}_{k_2,m_2}^{up}[n]$ and ${\bar R}_{k_1,m_2}^{up}[n]$, we have
\begin{align}
&({\rm P}2.1)\mathop {\max }\limits_{{P},{\tau _{{k_2}}}\left[ n \right],{\gamma _{ {k_2}}}\left[ n \right]} \sum\limits_{{k_2} = 1}^{{K_2}} {\sum\limits_{n = 1}^N {{\tau _{{k_2}}}\left[ n \right]} }\notag\\
{\rm s.t.}~&\sum\limits_{{m_2} = 1}^{{M_2}} {{x_{{k_2},{m_2}}}\left[ n \right]\left( {{{\bar R}_{{k_2},{m_2}}}\left[ n \right] - \tilde R_{{k_2},{m_2}}^{up}\left[ n \right]} \right)}  - {\gamma _{{k_2}}}\left[ n \right] \notag\\
&\qquad\qquad\qquad\qquad\qquad\quad\ge {\tau _{{k_2}}}\left[ n \right],n \in  {\cal N},{k_2},\\
&{\gamma _{{k_2}}}\left[ n \right] \ge \sum\limits_{{m_2} = 1}^{{M_2}} {{x_{{k_2},{m_2}}}\left[ n \right]\left( {\bar R_{{k_1},{m_2}}^{up}\left[ n \right] - {{\tilde R}_{{k_1},{m_2}}}\left[ n \right]} \right)} ,\notag\\
&\qquad\qquad\qquad\qquad\qquad\qquad\qquad,n \in  {\cal N},{k_1},{k_2},\\
&0 \le {p_i} \le {P_{\max }},i \in {{\cal M}_1} \cup {{\cal M}_2}.
\end{align}
$({\rm P}2.1)$ is a convex optimization problem, which can be efficiently solved by standard convex techniques. Consequently,  problem $({\rm P2})$ can be approximately solved by successively updating the transmit power based on the results obtained from $({\rm P}2.1)$.
\subsection{UAV trajectory optimization }
In this subsection, we consider the third sub-problem of $(\rm P)$ for optimizing UAV trajectory with given  transmit power and user scheduling. The SCA technique and Dinkelbach method are used to solve this non-convex sub-problem. Furthermore, the convex form does not obey the disciplined  CVX rules \cite{grant2014cvx}. To solve this problem, we equivalently reformulate it into recognized  rules without loss  of  optimality.

For the given UAV transmit power $P$ and user scheduling $X$, the third sub-problem of $(\rm P)$ is simplified as
\begin{align}
&({\rm{P}}3)\mathop {\max }\limits_{\scriptstyle Q,{\Phi _{{k_2}[n]}},\hfill\atop
\scriptstyle{\varphi _{{k_2}}}\left[ n \right]\hfill} \frac{{\sum\limits_{{k_2} = 1}^{{K_2}} {\sum\limits_{n = 1}^N {{\Phi _{{k_2}}}\left[ n \right]} } }}{{\sum\limits_{i \in {{\cal M}_1} \cup {{\cal M}_2}} {\sum\limits_{n = 1}^N {{c_1}{{\left\| {{{\bf{v}}_i}\left[ n \right]} \right\|}^3} + \frac{{{c_2}}}{{\left\| {{{\bf{v}}_i}\left[ n \right]} \right\|}}\left( {1 + \frac{{{{\left\| {{{\bf{a}}_i}\left[ n \right]} \right\|}^2}}}{{{g^2}}}} \right)} } }} \notag\\
&{\rm{s}}.{\rm{t}}.~\eqref{trajectory1}\text{-}\eqref{trajectory6},\notag\\
&\qquad{R_{{k_2}}}\left[ n \right] - {\varphi _{{k_2}}}\left[ n \right] \ge {\Phi _{{k_2}}}\left[ n \right],n \in  {\cal N},{k_2}, \label{p3const1}\\
&\qquad{\varphi _{{k_2}}}\left[ n \right] \ge {R_{{k_1} \to {k_2}}},n \in  {\cal N},{k_1},{k_2},\label{p3const2}
\end{align}
where $\Phi_{k_2}[n]$ and $\varphi_{k_2}$ are the slack variables. Note that the problem $(\rm P3)$ is non-convex due to the following reasons. First, the UAV trajectory $Q$ involved in \eqref{p3const1} and \eqref{p3const2} leads to  a  non-convex constraint set. Second, the UAV velocity  ${\bf v}_i[n]$ and acceleration ${\bf a}_i[n]$ are coupled in the objective function.  Therefore, it is intractable to directly solve the problem $(\rm P3)$ efficiently. In the following, we apply SCA  technique and Dinkelbach method to obtain the UAV trajectory. We first reformulate $(\rm P3)$ by introducing slack variables $\{ \mu_i[n]\}$ as
\begin{align}
&({\rm{P}}3.1)\mathop {\max }\limits_{\scriptstyle Q,{\Phi _{{k_2}}}\left[ n \right],\hfill\atop
\scriptstyle{\mu _i}\left[ n \right],{\varphi _{{k_2}}}\left[ n \right]\hfill} \frac{{\sum\limits_{{k_2} = 1}^{{K_2}} {\sum\limits_{n = 1}^N {{\Phi _{{k_2}}}\left[ n \right]} } }}{{\sum\limits_{i \in {{\cal M}_1} \cup {{\cal M}_2}} {\sum\limits_{n = 1}^N {{c_1}{{\left\| {{{\bf{v}}_i}\left[ n \right]} \right\|}^3} + \frac{{{c_2}}}{{{\mu _i}\left[ n \right]}} + \frac{{{c_2}{{\left\| {{{\bf{a}}_i}\left[ n \right]} \right\|}^2}}}{{{\mu _i}\left[ n \right]{g^2}}}} } }}\notag\\
&{\rm s.t.}~\eqref{trajectory1}\text{-}\eqref{trajectory6},\eqref{p3const1},\eqref{p3const2},\notag\\
&\qquad{\mu _i}\left[ n \right] \ge 0,n \in  {\cal N},i \in {{\cal M}_1} \cup {{\cal M}_2},\label{p3const3}\\
&\qquad{\left\| {{{\bf{v}}_i}\left[ n \right]} \right\|^2} \ge {\mu _i}\left[ n \right],n \in  {\cal N},i \in {{\cal M}_1} \cup {{\cal M}_2}\label{p3const4},
\end{align}
It can be verified  that at the optimal solution to  the problem $(\rm P3.1)$, we must have ${\left\| {{{\bf{v}}_i}\left[ n \right]} \right\|^2} = {\mu _i}\left[ n \right],n \in  {\cal N},i \in {{\cal M}_1} \cup {{\cal M}_2}$, which means the problem $(\rm P3)$ is equivalent to $(\rm P3.1)$. This can be proved by using reduction to absurdity. Suppose that  at the optimal value to  problem $(\rm P3.1)$, ${\left\| {{{\bf{v}}_i}\left[ n \right]} \right\|^2} > {\mu _i}\left[ n \right],n \in  {\cal N},i \in {{\cal M}_1} \cup {{\cal M}_2}$, one can always appropriately increase $\mu_i[n]$ to obtain a strictly larger objective value, which is contradictory to the assumption.  Therefore,  problem $(\rm P3.1)$ is equivalent to $(\rm P3)$.  Obviously, \eqref{p3const4} is a non-convex constraint. By applying the first-order Taylor expansion at local point $v_i^r[n]$, we have
\begin{align}
{\left\| {{v_i}\left[ n \right]} \right\|^2} \ge& {\left\| {v_i^r\left[ n \right]} \right\|^2} + 2{\left( {v_i^r\left[ n \right]} \right)^T}\left( {{v_i}\left[ n \right] - v_i^r\left[ n \right]} \right)\notag\\
& = {\Upsilon ^{lb}}\left( {{v_i}\left[ n \right]} \right),n \in  {\cal N},i \in {{\cal M}_1} \cup {{\cal M}_2},
\end{align}
Subsequently, the constraint \eqref{p3const4} can be rewritten as
\begin{align}
{\Upsilon ^{lb}}\left( {{v_i}\left[ n \right]} \right) \ge {\mu _i}\left[ n \right], n \in  {\cal N},i \in {{\cal M}_1} \cup {{\cal M}_2}, \label{p3const4NEW}
\end{align}
which is convex since ${\Upsilon ^{lb}}\left( {{v_i}\left[ n \right]} \right)$ is linear w.r.t. $v_i[n]$.

To tackle the non-convex constraint  \eqref{p3const1}, the SAC technique is leveraged. Specifically,  by introducing auxiliary variables $\{S_{k_2,i}[n],n \in  {\cal N},k_2 \in{\cal K}_2,i \in {\cal M}_1 \cup {\cal M}_2\}$ into ${\tilde R}_{k_2,m_2}[n]$ in  \eqref{p2_1},  ${\tilde R}_{k_2,m_2}[n]$ can be substituted as
\begin{align}
&{{\tilde{\tilde R}}_{{k_2,m_2}}}[n] = B{\log _2}\left( {\sum\limits_{i \in {{\cal M}_2}\backslash \left\{ {{m_2}} \right\} \cup {{\cal M}_1}} {\frac{{{p_i}\left[ n \right]{\beta _0}}}{{{H^2} + {S_{{k_2},i}}\left[ n \right]}}}  + {\sigma ^2}} \right),\notag\\
&\qquad\qquad\qquad\qquad\qquad\qquad \forall{k_2}, n \in {\cal N},m \in {\cal M}_2, \label{p3_1}
\end{align}
with additional constraint
\begin{align}
{S_{{k_2},i}}\left[ n \right] \le {\left\| {{{\bf{q}}_i}[n] - {{\bf{w}}_{{k_2}}}} \right\|^2},\forall {k_2},n\in{\cal N},i \in {{\cal M}_2}\backslash \left\{ {{m_2}} \right\} \cup {{\cal M}_1},\label{p3_1_0}
\end{align}
It is observed  from \eqref{p3_1_0} that  $\|{\bf q}_i[n]-{\bf w}_{k_2}[n]\|^2$  leads to a non-convex constraint set. To handle the non-convexity of \eqref{p3_1_0}, we have the following inequality by applying the first-order Taylor expansion at the given local point $\|{\bf q}_i^{r}[n]-{\bf w}_{k_2}\|$ at $r$\text{-th} iteration
\begin{align}
{\left\| {{{\bf{q}}_i}[n] - {{\bf{w}}_{{k_2}}}} \right\|^2} \ge& {\left\| {{\bf{q}}_i^r[n] - {{\bf{w}}_{{k_2}}}} \right\|^2} + 2{\left( {{\bf{q}}_i^r[n] - {{\bf{w}}_{{k_2}}}} \right)^T}\notag\\
&\times \left( {{{\bf{q}}_i}[n] - {\bf{q}}_i^r[n]} \right) = {\chi ^{lb}}\left( {{{\bf{q}}_i}[n]} \right),\label{p3_1_1}
\end{align}
which is convex since ${\chi ^{lb}}({\bf q}_i[n])$ is linear w.r.t. ${\bf q}_i[n]$.

Although ${\tilde{\tilde R}}$ is convex w.r.t.  $S_{k_2,i}[n]$ in \eqref{p3_1},  it does not obey the disciplined CVX  rules.
\begin{lemma}\label{lemma4}
To make the problem $(\rm P3.1)$ efficiently solved by CVX,  \eqref{p3_1} can be transformed  into
\begin{align}
{{\tilde {\tilde {\tilde R}}}_{{k_2,m_2}}}[n] = B\varpi[n] , \label{p3_2}
\end{align}
with additional constraints
\begin{align}
\left\{ \begin{array}{l}
\frac{{{H^2} + {S_{{k_2},i}}\left[ n \right]}}{{{p_i}\left[ n \right]{\beta _0}}} \ge {e^{ - {z_{{k_2},i}}\left[ n \right]}},\forall {k_2},i \in {{\cal M}_2}\backslash \left\{ {{m_2}} \right\} \cup {\cal M},n \in {\cal N},\\
\varpi[n]  \ge {\log _2}\left( {\sum\limits_{i \in {{\cal M}_2}\backslash \left\{ {{m_2}} \right\} \cup {{\cal M}_1}} {{e^{  {z_{{k_2,i}}}\left[ n \right]}}} }+\sigma^2 \right),
\end{array} \right. \label{p3_3}
\end{align}
\end{lemma}
where $z_{k_2,i[n]}$ and $\varpi[n]$ are the auxiliary variables. It can be seen that both \eqref{p3_2} and \eqref{p3_3} are  convex. However, ${\bar R}_{k_2,m_2}[n]$ in \eqref{p2_1} is not convex w.r.t. UAV trajectory ${\bf q}_i[n]$ but convex w.r.t. $\|{\bf q}_i[n]-{\bf w}_{k_2}[n]\|^2$. Thus, we have the following result.
\begin{lemma}\label{lemma5}
 With local point $\|{\bf q}_i^{r}[n]-{\bf w}_{k_2}[n]\|^2$, ${i \in {{\cal M}_1} \cup {{\cal M}_2}}$, $\forall k_2,$ over $r\text{-th}$ iteration, the inequality is hold in \eqref{p3_4} (at the top of next page).
\end{lemma}
\begin{IEEEproof}
The proof is similar to Lemma \ref{lemma2} and omitted here for simplicity.
\end{IEEEproof}
\newcounter{mytempeqncnt3}
\begin{figure*}
\normalsize
\setcounter{mytempeqncnt3}{\value{equation}}
\begin{align}
{{\bar R}_{{k_2},{m_2}}}\left[ n \right] \ge& B{\log _2}\left( {\sum\limits_{i \in {{\cal M}_1} \cup {{\cal M}_2}} {\frac{{{p_i}\left[ n \right]{\beta _0}}}{{{H^2} + {{\left\| {{\bf{q}}_i^r\left[ n \right] - {{\bf{w}}_{{k_2}}}} \right\|}^2}}}} } \right)-\notag\\
&B\sum\limits_{i \in {{\cal M}_1} \cup {{\cal M}_2}} {\frac{{\frac{{{p_i}\left[ n \right]{\beta _0}}}{{{{\left( {{H^2} + {{\left\| {{\bf{q}}_i^r\left[ n \right] - {{\bf{w}}_{{k_2}}}} \right\|}^2}} \right)}^2}}}{{\log }_2}\left( e \right)}}{{\sum\limits_{j \in {{\cal M}_1} \cup {{\cal M}_2}} {\frac{{{p_j}\left[ n \right]{\beta _0}}}{{{H^2} + {{\left\| {{\bf{q}}_j^r\left[ n \right] - {{\bf{w}}_{{k_2}}}} \right\|}^2}}}} }}\left( {{{\left\| {{{\bf{q}}_i}\left[ n \right] - {{\bf{w}}_{{k_2}}}} \right\|}^2} - {{\left\| {{\bf{q}}_i^r\left[ n \right] - {{\bf{w}}_{{k_2}}}} \right\|}^2}} \right)}  \overset{\triangle}= \bar R_{{k_2},{m_2}}^{lb}\left[ n \right].\label{p3_4}
\end{align}
\hrulefill 
\vspace*{4pt} 
\end{figure*}

Based on above transformations, the constraint \eqref{p3const1} can be reformulated as
\begin{align}
\sum\limits_{{m_2} = 1}^{{M_2}} {{x_{{k_2},{m_2}}}\left[ n \right]\left( {\bar R_{{k_2},{m_2}}^{lb}\left[ n \right] - {{{\tilde {\tilde {\tilde R}}}}_{{k_2,i}}}[n]} \right)}  - {\varphi _{{k_2}}}\left[ n \right]\notag\\
\ge {\Phi _{{k_2}}}\left[ n \right],\forall n,{k_2}.\label{p3const1NEW}
\end{align}
Next, by introducing auxiliary variables $\{Z_{k_1,i}[n],n \in  {\cal N},\forall k_1,i \in {\cal M}_1 \cup {\cal M}_2\}$ into ${\bar R}_{k_1,m_2}[n]$ in \eqref{p2_2}, we deal with the non-convex constraint  \eqref{p3const2}, which can be substituted as
\begin{align}
{{{\bar {\bar R}}}_{{k_1,m_2}}}[n] = B{\log _2}\left( {\sum\limits_{i \in {{\cal M}_1} \cup {{\cal M}_2}} {\frac{{{p_i}\left[ n \right]{\beta _0}}}{{{H^2} + {Z_{k_1,i}}\left[ n \right]}}}  + {\sigma ^2}} \right),\notag\\
\forall k_1,n \in {\cal N}, \label{p3_5}
\end{align}
with additional constraint
\begin{align}
{Z_{{k_1},i}}\left[ n \right] \le {\left\| {{{\bf{q}}_i}[n] - {{\bf{w}}_{{k_1}}}} \right\|^2},\forall {k_1},n\in{N} ,i \in {{\cal M}_1} \cup {{\cal M}_2}.\label{p3_5_0}
\end{align}

For \eqref{p3_5_0},  by applying the first-order Taylor expansion at the given local point $\|{\bf q}_i^{r}[n]-{\bf w}_{k_1}\|$ over $r$\text{-th} iteration, we have the following inequality
\begin{align}
{\left\| {{{\bf{q}}_i}[n] - {{\bf{w}}_{{k_1}}}} \right\|^2} \ge& {\left\| {{\bf{q}}_i^r[n] - {{\bf{w}}_{{k_1}}}} \right\|^2} + 2{\left( {{\bf{q}}_i^r[n] - {{\bf{w}}_{{k_1}}}} \right)^T}\notag\\
&\times \left( {{{\bf{q}}_i}[n] - {\bf{q}}_i^r[n]} \right) = {\Xi  ^{lb}}\left( {{{\bf{q}}_i}[n]} \right).\label{p3_5_1}
\end{align}
Note  that \eqref{p3_5} has same structure with \eqref{p3_1}, it  can be reformulated it as  following disciplined CVX  form
\begin{align}
{{\bar {\bar {\bar R}}}_{{k_1,m_2}}}[n] = B{\bar \varpi}[n] , \label{p3_6}
\end{align}
with additional constraints
\begin{align}
\left\{ \begin{array}{l}
\frac{{{H^2} + {Z_{{k_1},i}}\left[ n \right]}}{{{p_i}\left[ n \right]{\beta _0}}} \ge {e^{ - {{\bar z}_{{k_1},i}}\left[ n \right]}},\forall k_1, i \in {\cal M}_1 \cup {\cal M}_2,n \in {\cal N},\\
{\bar \varpi}[n]  \ge {\log _2}\left( {\sum\limits_{i \in {{\cal M}_1} \cup {{\cal M}_2}} {{e^{  {{\bar z}_{{k_1,i}}}\left[ n \right]}}} } +\sigma^2\right),\label{p3_7}
\end{array} \right.
\end{align}
where ${\bar z}_{k_1,i[n]}$ and ${\bar \varpi}[n]$ are the auxiliary variables. It is worth mentioning that in \eqref{p2_2}, ${\tilde R}_{k_1,m_2}$ is neither convex nor concave w.r.t. ${\bf q}_i[n]$ while it is convex w.r.t. $\|{\bf q}_i[n]-{\bf w}_{k_1}[n]\|^2$. Then, we can obtain the following lemma.
\begin{lemma}\label{lemma6}
 With given local point $\|{\bf q}_i^r[n]-{\bf w}_{k_1}[n]\|^2,i \in {{\cal M}_2}\backslash \left\{ {{m_2}} \right\} \cup {{\cal M}_1},$ over $r\text{-th}$ iteration, the inequality is hold in \eqref{p3_8} (at the top of next page).
\end{lemma}
\begin{IEEEproof}
The proof is similar to Lemma \ref{lemma2} and omitted here for simplicity.
\end{IEEEproof}
\newcounter{mytempeqncnt4}
\begin{figure*}
\normalsize
\setcounter{mytempeqncnt4}{\value{equation}}
\begin{align}
&{{\tilde R}_{{k_1},{m_2}}}\left[ n \right] \ge B{\log _2}\left( {\sum\limits_{i \in {{\cal M}_2}\backslash \left\{ {{m_2}} \right\} \cup {{\cal M}_1}} {\frac{{{p_i}\left[ n \right]{\beta _0}}}{{{H^2} + {{\left\| {{\bf{q}}_i^r\left[ n \right] - {{\bf{w}}_{{k_1}}}} \right\|}^2}}}} } \right)\notag\\
&\quad\quad\quad- B\sum\limits_{i \in {{\cal M}_2}\backslash \left\{ {{m_2}} \right\} \cup {{\cal M}_1}} {\frac{{\frac{{{p_i}\left[ n \right]{\beta _0}}}{{{{\left( {{H^2} + {{\left\| {{\bf{q}}_i^r\left[ n \right] - {{\bf{w}}_{{k_1}}}} \right\|}^2}} \right)}^2}}}{{\log }_2}\left( e \right)}}{{\sum\limits_{j \in {{\cal M}_2}\backslash \left\{ {{m_2}} \right\} \cup {{\cal M}_1}} {\frac{{{p_j}\left[ n \right]{\beta _0}}}{{{H^2} + {{\left\| {{\bf{q}}_j^r\left[ n \right] - {{\bf{w}}_{{k_1}}}} \right\|}^2}}}} }}\left( {{{\left\| {{{\bf{q}}_i}\left[ n \right] - {{\bf{w}}_{{k_1}}}} \right\|}^2} - {{\left\| {{\bf{q}}_i^r\left[ n \right] - {{\bf{w}}_{{k_1}}}} \right\|}^2}} \right)} \overset{\triangle} = \tilde R_{{k_1},{m_2}}^{lb}\left[ n \right].\label{p3_8}
\end{align}
\hrulefill 
\vspace*{4pt} 
\end{figure*}

Then, the constraint \eqref{p3const2} can be rewritten as
\begin{align}
&{\varphi _{{k_2}}}\left[ n \right] \ge \sum\limits_{{m_2} = 1}^{{M_2}} {{x_{{k_2},{m_2}}}\left[ n \right]\left( {{{{\bar {\bar {\bar R}}}}_{{k_1},{m_2}}}\left[ n \right] - \tilde R_{{k_1},{m_2}}^{lb}\left[ n \right]} \right)},\notag\\
&\qquad\qquad \qquad \qquad \qquad \qquad \qquad \qquad  n \in {\cal N},\forall {k_1},{k_2},\label{p3const2NEW}
\end{align}

By defining the auxiliary variables set $\Theta  = \{\left( {{\mu _i}\left[ n \right],{S_{{k_2},i}}\left[ n \right],\varpi ,{z_{{k_2},i}}\left[ n \right],{Z_{{k_1},i}}\left[ n \right],\bar \varpi ,{{\bar z}_{{k_1},i}}\left[ n \right]} \right)\}$, the problem $(\rm P3)$ can be simplified as
\begin{align}
&({\rm{P}}3.2)\mathop {\max }\limits_{\scriptstyle Q,{\Phi _{{k_2}}}[n],\hfill\atop
\scriptstyle\Theta ,{\varphi _{{k_2}}}\left[ n \right]\hfill} \frac{{\sum\limits_{{k_2} = 1}^{{K_2}} {\sum\limits_{n = 1}^N {{\Phi _{{k_2}}}\left[ n \right]} } }}{{\sum\limits_{i \in {{\cal M}_1} \cup {{\cal M}_2}} {\sum\limits_{n = 1}^N {{c_1}{{\left\| {{{\bf{v}}_i}\left[ n \right]} \right\|}^3} + \frac{{{c_2}}}{{{\mu _i}\left[ n \right]}} + \frac{{{c_2}{{\left\| {{{\bf{a}}_i}\left[ n \right]} \right\|}^2}}}{{{\mu _i}\left[ n \right]{g^2}}}} } }}\notag\\
&{\rm s.t.}~\eqref{trajectory1}\text{-}\eqref{trajectory6},\eqref{p3const3}, \eqref{p3const4NEW}, \eqref{p3_3},\eqref{p3const1NEW},\eqref{p3_7},\eqref{p3const2NEW}\notag\\
&{S_{{k_2},i}}\left[ n \right] \le {\chi ^{lb}}\left( {{{\bf{q}}_i}[n]} \right),\forall {k_2},n \in N,i \in {{\cal M}_2}\backslash \left\{ {{m_2}} \right\} \cup {{\cal M}_1},\notag\\
&{Z_{{k_1},i}}\left[ n \right] \le {\Xi ^{lb}}\left( {{{\bf{q}}_i}[n]} \right),\forall {k_1},n \in N,i \in {{\cal M}_1} \cup {{\cal M}_2}.\notag
\end{align}

It can be seen that  problem  $({\rm P3.2})$ is a fractional maximization problem  with  convex constraints,   convex  denominator and linear numerator, wherein Dinkelbach method can be employed. For the analytic simplicity,  define $\cal F$ as the set of feasible points of problem $({\rm P3.2})$, and  ${{\tilde E}_{total}} = \sum\limits_{i \in {{\cal M}_1} \cup {{\cal M}_2}} {\sum\limits_{n = 1}^N {{c_1}{{\left\| {{{\bf{v}}_i}\left[ n \right]} \right\|}^3} + \frac{{{c_2}}}{{{\mu _i}\left[ n \right]}} + \frac{{{{\left\| {{{\bf{a}}_i}\left[ n \right]} \right\|}^2}}}{{{\mu _i}\left[ n \right]{g^2}}}} }$.  In addition, by defining $\zeta^\ast $  as the maximum security energy efficiency, we have
\begin{align}
{\zeta ^ \ast } = \mathop {\max }\limits_{Q,{\Phi _{{k_2}}}\left[ n \right],{\Theta},{\varphi _{{k_2}}}\left[ n \right]} \frac{{\sum\limits_{{k_2} = 1}^{{K_2}} {\sum\limits_{n = 1}^N {{\Phi _{{k_2}}}\left[ n \right]} } }}{{{{\tilde E}_{total}}}}. \label{p3_9}
\end{align}
Hence, we have the following theorem.
\begin{theorem}\label{theorem1}
The optimal solutions of problem $({\rm P3.2})$ achieve the  maximum SEE value $\zeta^\ast $ if and only if
\begin{align}
\mathop {\max }\limits_{Q,{\Phi _{{k_2}}}\left[ n \right],{\Theta},{\varphi _{{k_2}}}\left[ n \right]} \sum\limits_{{k_2} = 1}^{{K_2}} {\sum\limits_{n = 1}^N {{\Phi _{{k_2}}}\left[ n \right]} }  - {\zeta^\ast }{{\tilde E}_{total}} = 0.\label{p3_10}
\end{align}
\end{theorem}
\begin{IEEEproof}
The proof can be referred to \cite{dinkelbach1967nonlinear,zappone2017Globally,ng2012energy}.
\end{IEEEproof}

Then, we can rewrite the problem $({\rm P3.2})$ as
\begin{align}
&({\rm P3.3})\mathop {\max }\limits_{Q,{\Phi _{{k_2}}}\left[ n \right],{\Theta},{\varphi _{{k_2}}}\left[ n \right]} \sum\limits_{{k_2} = 1}^{{K_2}} {\sum\limits_{n = 1}^N {{\Phi _{{k_2}}}\left[ n \right]} }  - {\zeta }{{\tilde E}_{total}}\notag\\
&{\rm s.t.}~\eqref{trajectory1}\text{-}\eqref{trajectory6},\eqref{p3const3}, \eqref{p3const4NEW}, \eqref{p3_3},\eqref{p3const1NEW},\eqref{p3_7},\eqref{p3const2NEW},\notag\\
&{S_{{k_2},i}}\left[ n \right] \le {\chi ^{lb}}\left( {{{\bf{q}}_i}[n]} \right),\forall {k_2},n \in N,i \in {{\cal M}_2}\backslash \left\{ {{m_2}} \right\} \cup {{\cal M}_1},\notag\\
&{Z_{{k_1},i}}\left[ n \right] \le {\Xi ^{lb}}\left( {{{\bf{q}}_i}[n]} \right),\forall {k_1},n \in N,i \in {{\cal M}_1} \cup {{\cal M}_2}.\notag
\end{align}
Finally, an iterative algorithm, namely Dinkelbach method \cite{dinkelbach1967nonlinear}, is proposed for solving problem $({\rm P3.2})$, which is summarized in Algorithm 2. It is worth pointing out that the global optimality can be guaranteed by using Dinkelbach method. As a consequence, the  problem $({\rm P3.2})$ can be  optimally solved by using Algorithm 2, and problem $({\rm P3})$ can be approximately  solved by successively updating the UAV trajectory based on the optimal solution to problem $({\rm P3.2})$.
\begin{algorithm}[H]
\caption{Dinkelbach method for solving  problem $(\rm P3.2)$}
\label{alg1}
\begin{algorithmic}[1]
\STATE  \textbf{Initialization:}  the maximum tolerance $\epsilon$, $\zeta=0 $ and iteration index $t$.
\STATE  \textbf{Repeat}
\STATE  \quad Solve problem $({\rm P3.3})$ with a given $\zeta$, denote $\{Q^ \ast,$ \\
\quad ${{\Phi^ \ast _{{k_2}}}\left[ n \right],{\Theta^ \ast},{\varphi^ \ast _{{k_2}}}\left[ n \right]}\}$ as obtained optimal solutions.
\STATE  \quad \textbf{if}\quad $\sum\limits_{{k_2} = 1}^{{K_2}} {\sum\limits_{n = 1}^N {{\Phi^ \ast _{{k_2}}}\left[ n \right]} }  - \zeta {{\tilde E}^ \ast_{total}} \le \epsilon$ \textbf{then}\\
\STATE  \qquad Convergence=\textbf{true}.   \\
\STATE \qquad  \textbf{Output} $\left\{ {Q^ \ast,{\Phi^ \ast _{{k_2}}}\left[ n \right],{\Theta^ \ast},{\varphi^ \ast _{{k_2}}}\left[ n \right]} \right\}$ and
\\ \qquad ${\zeta ^ \ast } = \frac{{\sum\limits_{{k_2} = 1}^{{K_2}} {\sum\limits_{n = 1}^N {{\Phi^\ast _{{k_2}}}\left[ n \right]} } }}{{{{\tilde E}^\ast_{total}}}}$.\\
\STATE  \quad  \textbf{else} \\
\STATE  \qquad Set $\zeta ^ \ast = \frac{{\sum\limits_{{k_2} = 1}^{{K_2}} {\sum\limits_{n = 1}^N {{\Phi _{{k_2}}}\left[ n \right]} } }}{{{{\tilde E}_{total}}}}$ and $t=t+1$.\\
\STATE  \qquad $\zeta=\zeta ^ \ast$.\\
\STATE  \qquad  Convergence=\textbf{false}.   \\
 \STATE  \quad \textbf{end if}\\
\STATE   \textbf{Until} Convergence=\textbf{true}.
\end{algorithmic}
\end{algorithm}
\subsection{Overall iterative algorithm }
In this subsection, we solve problem $(\rm P)$ by using the  block coordinate descent method \cite{luo1992convergence}. Specifically, in the $l$\text{-th} iteration, we first  obtain the optimal  user scheduling $X^\ast_{l+1}$ with any given transmit power $P_l$ and UAV trajectory $Q_l$ by solving problem $(\rm P1)$. Next, the optimal transmit power  $P^\ast_{l+1}$  with any given   user scheduling $X^\ast_{l+1}$ and UAV trajectory $Q_l$ is obtained  by solving problem $(\rm P2)$. Then,  the optimal  UAV trajectory $Q^\ast_{l+1}$ with any given   user scheduling $X^\ast_{l+1}$ and  transmit power  $P^\ast_{l+1}$ is obtained  by solving problem $(\rm P3)$. The details of this algorithm are summarized in Algorithm 3.
\begin{algorithm}[H]
\caption{Block coordinate descent method for solving  problem $(\rm P)$}
\label{alg2}
\begin{algorithmic}[1]
\STATE  \textbf{Initialization:}  the maximum tolerance $\epsilon$, UAV trajectory $Q_0$,  transmit power $P_0$ and iteration index $l$.
\STATE  \textbf{Repeat}
\STATE  \quad Solve problem $(\rm P1)$ for any given transmit power $P_l$\\
\quad    and UAV trajectory $Q_l$, and denote the obtained optimal \\
\quad  user  scheduling  as $X^\ast_{l+1}$.
\STATE  \quad Solve problem $(\rm P2)$ for any given  user scheduling $X^\ast_{l+1}$\\
\quad  and  UAV trajectory $Q_l$, and denote the obtained  optimal\\
\quad  transmit  power  as $P^\ast_{l+1}$.
\STATE  \quad Solve problem $(\rm P3)$ for any given user scheduling $X^\ast_{l+1}$ \\
\quad  and  transmit power $P^\ast_{l+1}$, and denote the obtained  UAV \\
\quad    trajectory as $Q^\ast_{l+1}$.
\STATE  Update $l=l+1$.
\STATE   \textbf{Until} the fractional increase of the objective value of $(\rm P)$ is less than tolerance  $\epsilon$.
\end{algorithmic}
\end{algorithm}
Next, a brief illustration of the convergence property of the Algorithm 3 is discussed. Let us define the objective value of problem $(\rm P1)$, $(\rm P2)$  and $(\rm P3)$ over $l$\text{-th} iteration as $R_1(X_l,P_l,Q_l)$, $R_2(X_l,P_l,Q_l)$ and $EE(X_l,P_l,Q_l)$, respectively. Then, at the $l+1$\text{-th} iteration, we have the relationships as \eqref{p4const1} (at the top of the next page). In \eqref{p4const1}, the inequalities $a$,  $c$ and $e$ follow  the fact that the optimal results are obtained by problem $(\rm P1)$, $(\rm P2)$  and $(\rm P3)$, respectively. The formula \eqref{p4const1} indicates that  Algorithm 3 is non-decreasing over each iteration. Apart from this, since the objective value of problem $(\rm P)$ is upper bounded by a finite value,  Algorithm 3 is thus  guaranteed to converge.

In Algorithm 3,  the computational complexity of solving $(\rm P1)$ is ${\cal O}\left( {N\left( {2{K_2}{M_2}  + {C_{\bar m_2^y}}{N_{\bar m_2^y}}} \right)} \right)$. For problem $(\rm P2)$, it involves logarithmic constraints, which can be approximated as linear constraints by taking first-order Taylor expansion. Then the problem becomes a linear problem and can be solved by interior point method with computational complexity ${\cal O}\left( {\sqrt {\left( {{M_1}{\rm{ + }}{M_2}} \right)N{\rm{ + }}2{K_2}N} \log \frac{1}{{{\varepsilon _0}}}} \right)$, where ${\left( {{M_1}{\rm{ + }}{M_2}} \right)N{\rm{ + }}2{K_2}N}$ denotes the decision variables, and $\varepsilon _0$ denotes iterative accuracy \cite{gondzio1996computational}. Similarly to  $(\rm P2)$, the computational complexity of  $(\rm P3)$ is ${\cal O}\left( {{L_1}\left( {\sqrt {4N + \left( {2{K_1} + 2{K_2} + N} \right)\left( {{M_1} + {M_2}} \right)} \log \frac{1}{{{\varepsilon _0}}}} \right)} \right)$, where $L_1$ denotes the iteration  numbers for updating $\zeta $ in Algorithm 2. Thus, the overall computational complexity of Algorithm 3 is  ${\cal O}\left( {{L_1}\left( {\sqrt {4N + \left( {2{K_1} + 2{K_2} + N} \right)\left( {{M_1} + {M_2}} \right)} \log \frac{1}{{{\varepsilon _0}}}} \right) + } \right.$ $\sqrt {\left( {{M_1}{\rm{ + }}{M_2}} \right)N{\rm{ + }}2{K_2}N} \log \frac{1}{{{\varepsilon _0}}}\left. { + N\left( {2{K_2}{M_2} + {C_{\bar m_2^y}}{N_{\bar m_2^y}}} \right)} \right)$.

\newcounter{mytempeqncnt5}
\begin{figure*}
\normalsize
\setcounter{mytempeqncnt5}{\value{equation}}
\begin{align}
EE\left( {{X_l},{P_l},{Q_l}} \right) = \frac{{{R_1}\left( {{X_l},{P_l},{Q_l}} \right)}}{{{E_{{\rm{total}}}}\left( {{Q_l}} \right)}} & \overset{a}\le \frac{{{R_1}\left( {{X_{l + 1}},{P_l},{Q_l}} \right)}}{{{E_{{\rm{total}}}}\left( {{Q_l}} \right)}} = \frac{{{R_2}\left( {{X_{l + 1}},{P_l},{Q_l}} \right)}}{{{E_{{\rm{total}}}}\left( {{Q_l}} \right)}} \notag\\
&\overset{b}\le \frac{{{R_2}\left( {{X_{l + 1}},{P_{l + 1}},{Q_l}} \right)}}{{{E_{{\rm{total}}}}\left( {{Q_l}} \right)}} = EE\left( {{X_{l + 1}},{P_l},{Q_l}} \right) \overset{c}\le EE\left( {{X_{l + 1}},{P_{l + 1}},{Q_{l + 1}}} \right).\label{p4const1}
\end{align}
\hrulefill 
\vspace*{4pt} 
\end{figure*}
\begin{table}
  \centering
  \caption{Simulation Parameters}\label{SimParams}
    \begin{tabular}{|c|c|}   
    \hline
    \textbf{Parameter} & \textbf{ Value} \\
    \hline
      Channel gain: ${\beta _0}$ &     $- 60{\rm{dB}}$ \cite{wu2018Joint}\\
    \hline
      Noise power: $\sigma^2 $ & $-110{\rm dBm}$ \cite{wu2018Joint}\\
    \hline
     Duration of each time slot: $\delta$ & $0.5{\rm s}$\cite{zeng2018energy}\\
    \hline
    System bandwidth: $B $ & $1\rm{ MHz}$ \cite{zeng2018energy}\\
     \hline
    UAV altitude: ${H }$ & $100\rm{m}$ \cite{zeng2018energy}\\
    \hline
    UAV maximum transmit power: $P_{\rm max}$ &  $1{\rm W}$ \cite{alzenad20173}\\
    \hline
    UAV maximum speed: $v_{\rm max }$ & $ 50{\rm{m/s}}$ \cite{wu2018Joint}\\
    \hline
    UAV maximum acceleration: $a_{\rm max }$ & $ 5{\rm{m/s^2}}$ \cite{zeng2018energy}\\
    \hline
    UAV energy consumption coefficient: $c_1$ & $9.26\times10^{-4}$ \cite{zeng2018energy} \\
    \hline
    UAV energy consumption coefficient: $c_2$ & $2250$ \cite{zeng2018energy}\\
    \hline
    Maximum tolerance: $\epsilon$ & $10^{-2}$\\
    \hline
    \end{tabular}
\end{table}

\begin{figure}[!t]
\centerline{\includegraphics[width=2.8in]{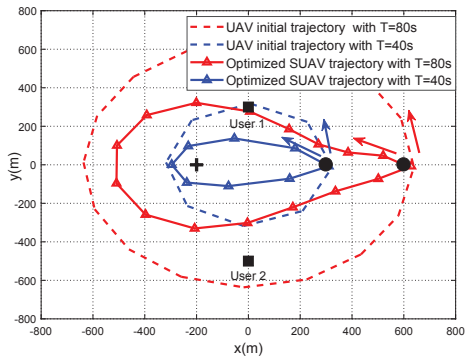}}
\caption{Optimized UAV trajectories for different period $T$ with a single SUAV. Each optimized trajectory is sampled every $5s$ and the sampled  points are marked with $\bigtriangleup$. The direction of arrow is the  direction of the initial trajectory. The legitimate users locations are marked by $\blacksquare $ and the eavesdropper location is marked by \arc. In addition, the initial location of UAV is marked by $\sbullet[1.5]$.} \label{fig2}
\end{figure}

%

\begin{figure}[!t]
\centerline{\includegraphics[width=2.8in]{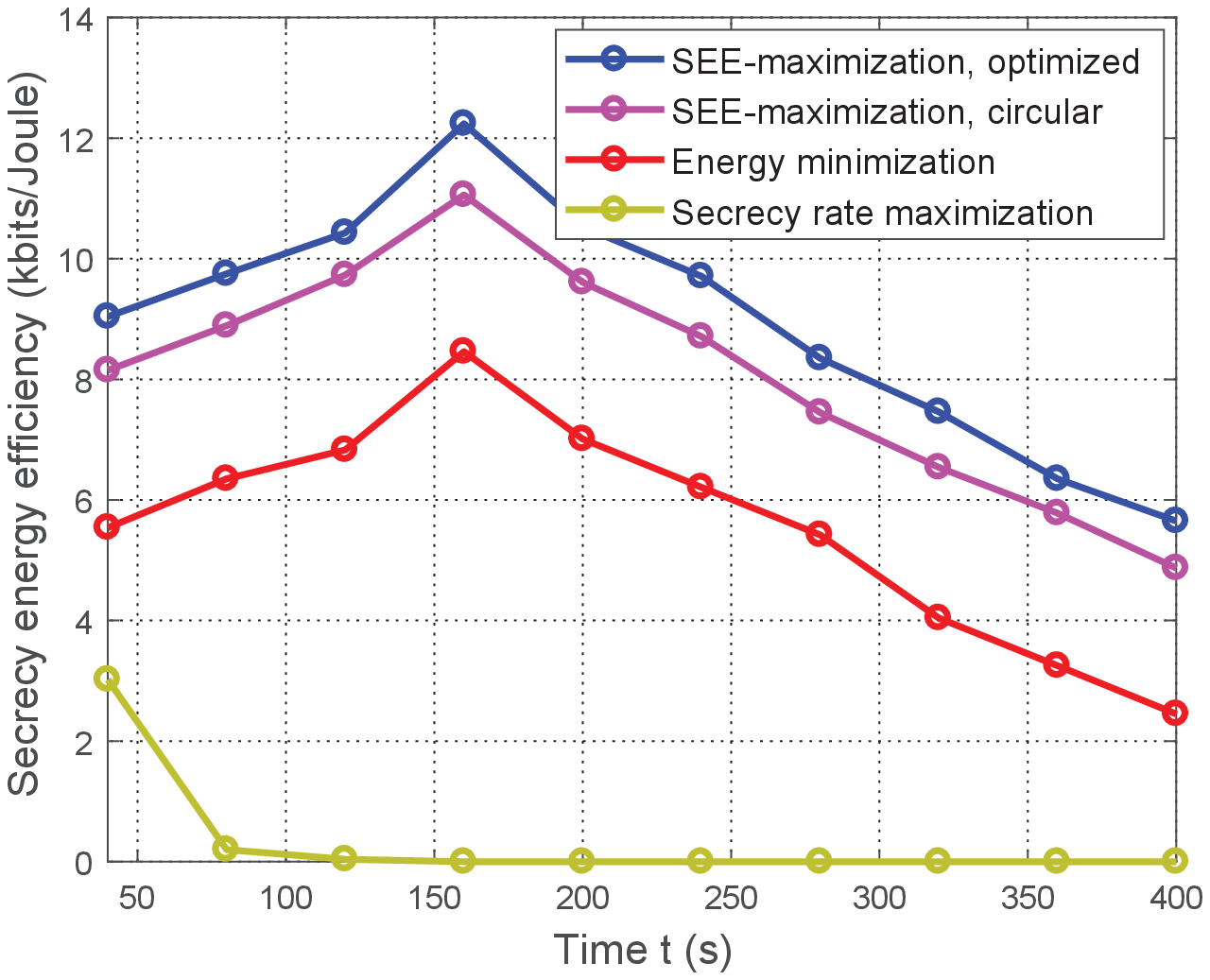}}
\caption{The SEE comparison of different schemes in scenario A.} \label{fig5}
\end{figure}

\begin{figure}[!t]
\centerline{\includegraphics[width=2.8in]{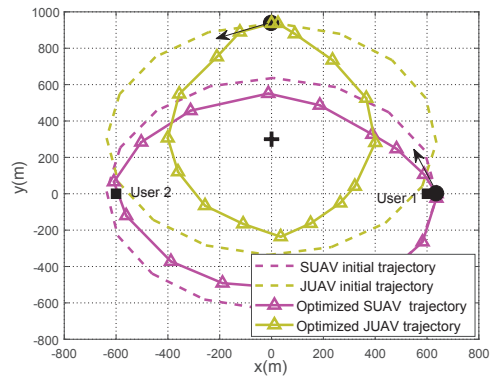}}
\caption{Optimized SUAV trajectory and JUAV trajectory for  period $T=80s$. Each optimized trajectory is sampled every $5s$ and the sampled  points are marked with $\bigtriangleup$. The direction of arrow is the  direction of the initial trajectory. The legitimate users locations are marked by $\blacksquare $ and the eavesdropper location is marked by \arc. In addition, the initial location of UAV is marked by $\sbullet[1.5]$.} \label{fig6}
\end{figure}

\begin{figure}[!t]
\centerline{\includegraphics[width=2.8in]{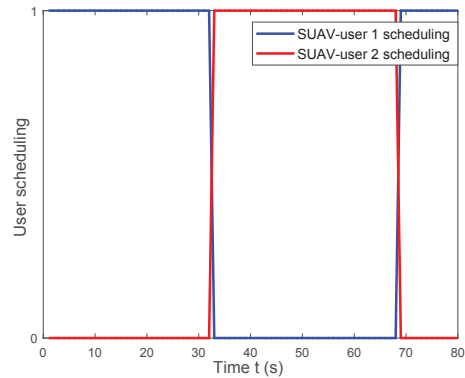}}
\caption{UAV-user scheduling for period $T=80s$. } \label{fig7}
\end{figure}

\begin{figure}[!t]
\centerline{\includegraphics[width=2.8in]{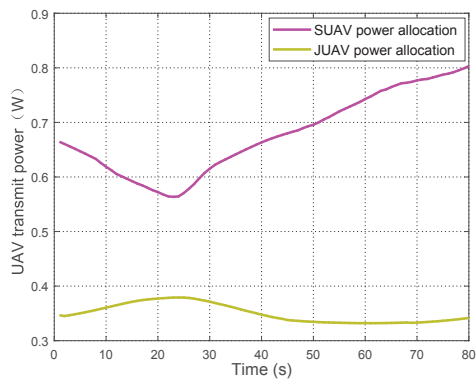}}
\caption{UAV transmit power  for period $T=80s$.} \label{fig8}
\end{figure}

\begin{figure}[!t]
\centerline{\includegraphics[width=2.8in]{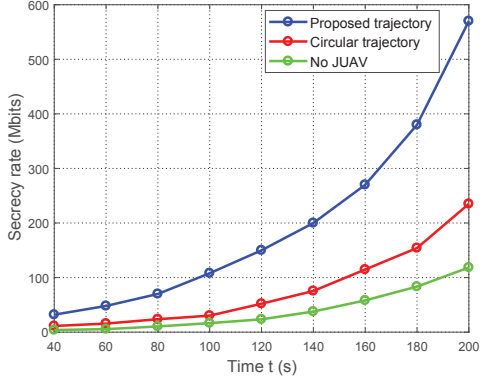}}
\caption{Secrecy rate performance versus different period $T$. } \label{fig9}
\end{figure}

\begin{figure}[htbp]
\centerline{\includegraphics[width=2.8in]{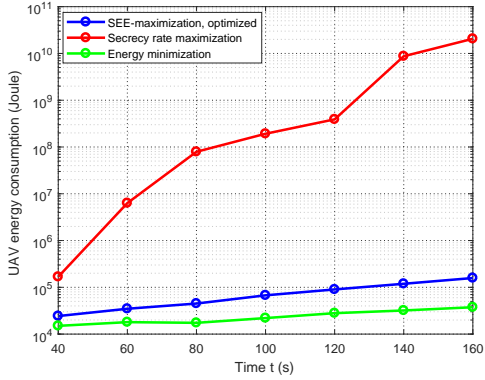}}
\caption{UAV energy consumption versus period $T$ for the One SUAV and One JUAV Case with different schemes. }\label{figtemp9}
\end{figure}

\begin{figure}[!t]
\centerline{\includegraphics[width=2.8in]{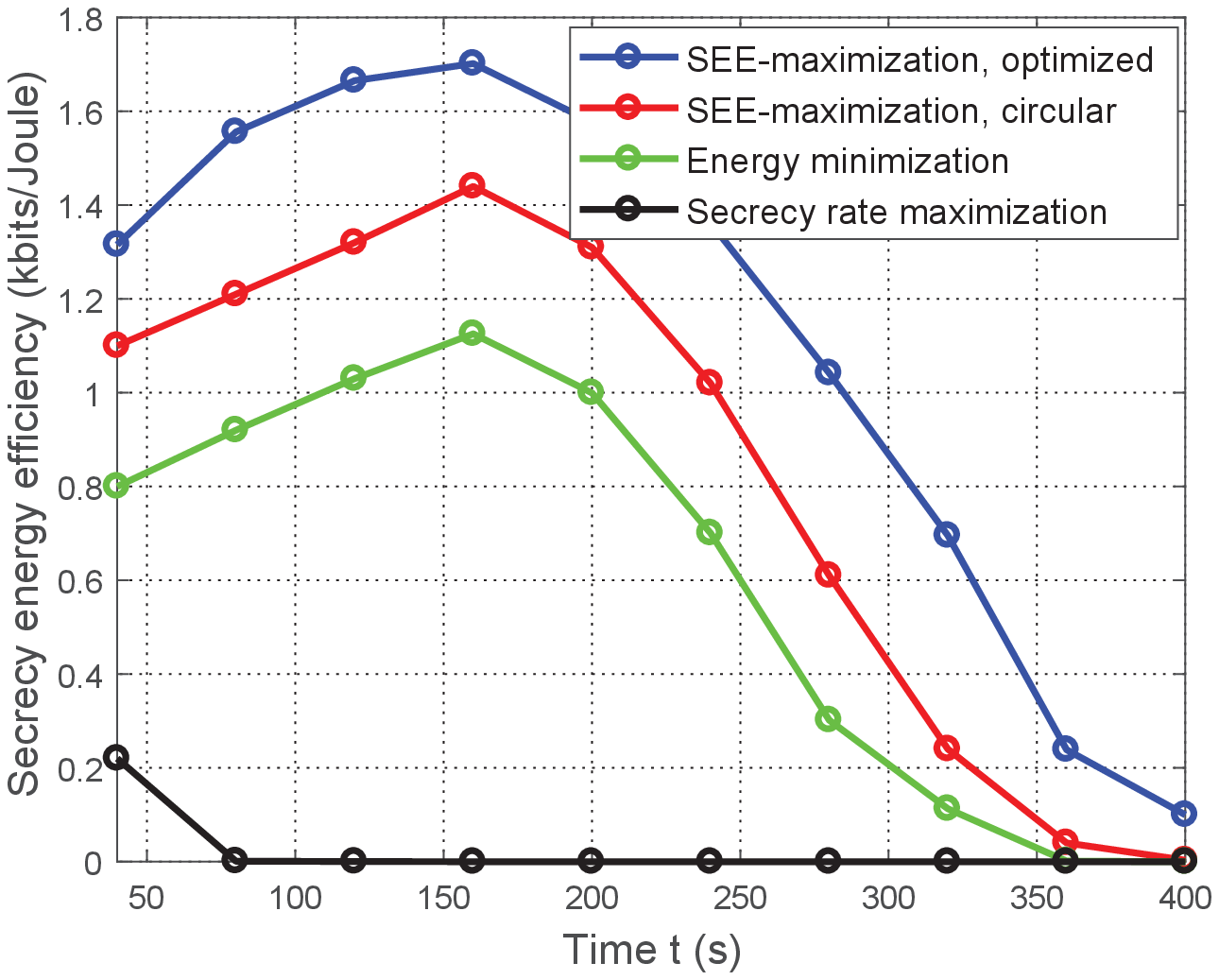}}
\caption{The SEE comparison of different schemes in scenario B.} \label{fig10}
\end{figure}
\begin{figure}[!t]
\centerline{\includegraphics[width=3in]{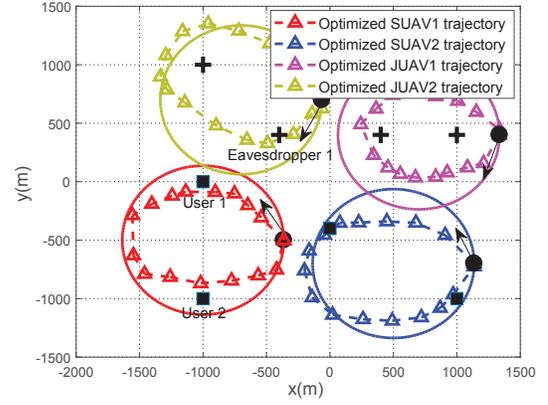}}
\caption{Optimized two SUAV trajectories and two JUAV trajectories for  period $T=80s$. Each optimized trajectory is sampled every $5s$ and the sampled  points are marked with $\bigtriangleup$. The solid circles are the UAV initial trajectories with same color corresponding its optimized UAV trajectories. The direction of arrow is the  direction of the initial trajectory. The legitimate users locations are marked by $\blacksquare $ and the eavesdropper location is marked by \arc. In addition, the initial location of UAV is marked by $\sbullet[1.5]$.} \label{fig11}
\end{figure}

\begin{figure}[!t]
\centerline{\includegraphics[width=3in]{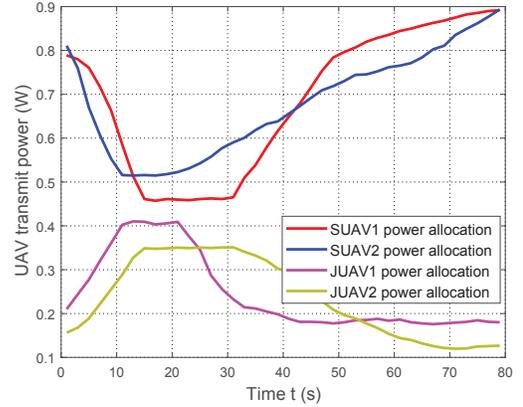}}
\caption{UAV transmit power with SUAV and JUAV for period $T=80s$.} \label{fig12}
\end{figure}
\begin{figure}[!t]
\centerline{\includegraphics[width=3in]{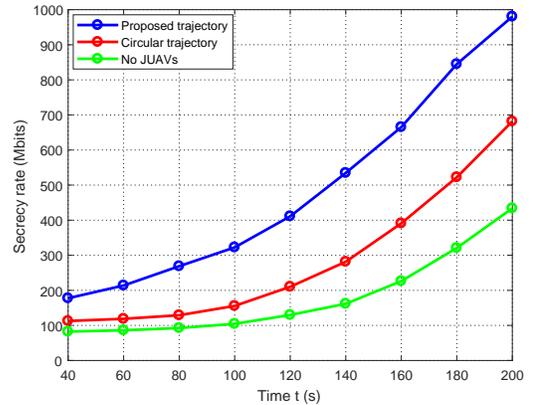}}
\caption{Secrecy rate performance of  two SUAVs and two JUAVs versus different period $T$. } \label{fig13}
\end{figure}
\begin{figure}[htbp]
\centerline{\includegraphics[width=2.8in]{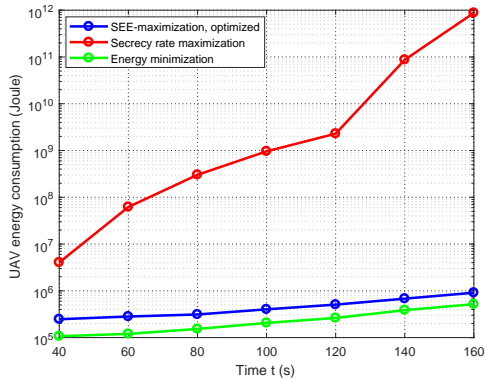}}
\caption{UAV energy consumption versus period $T$ for the Multi-SUAV and Multi-JUAV  Case with different schemes. }\label{figtemp13}
\end{figure}

\begin{figure}[!t]
\centerline{\includegraphics[width=3in]{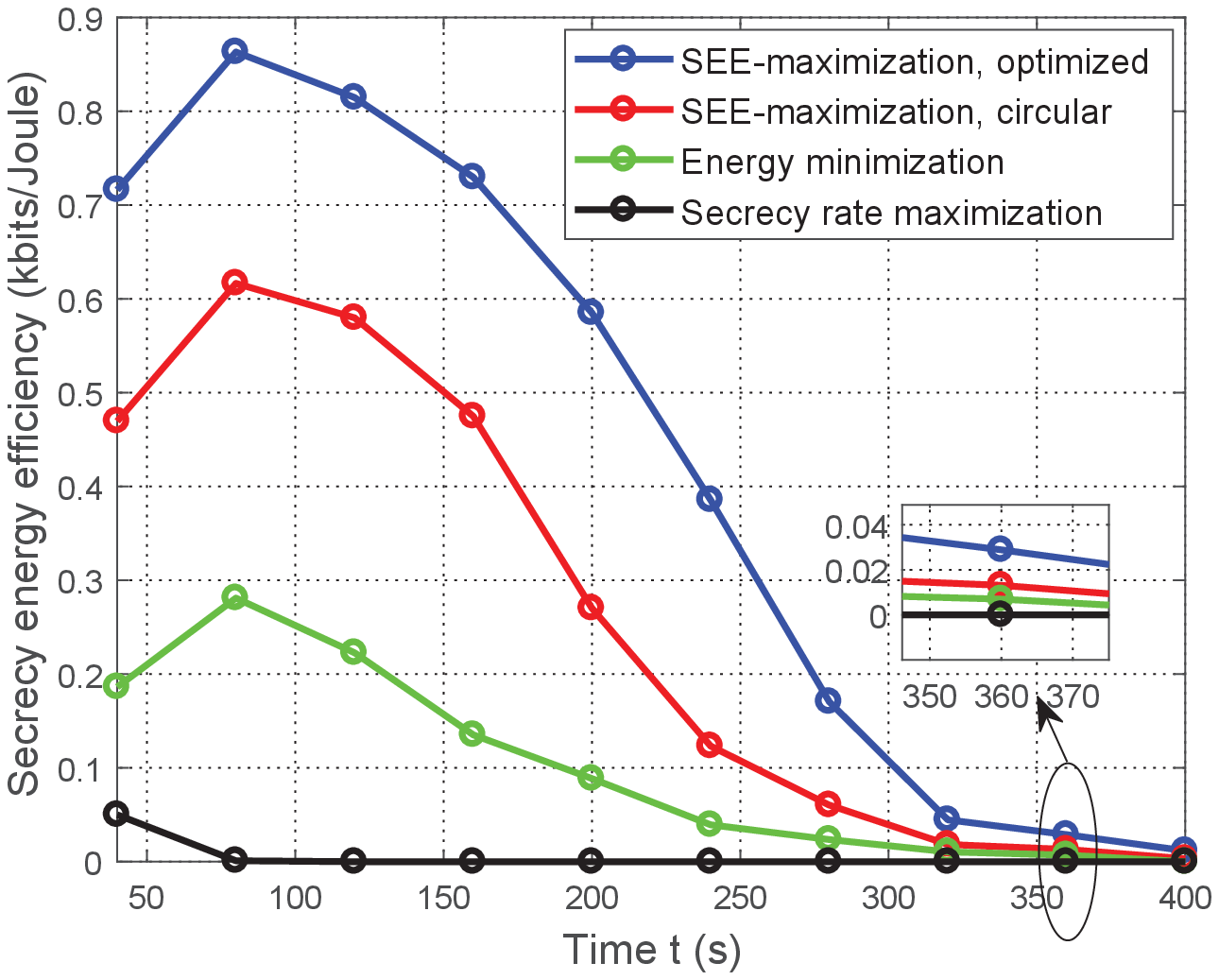}}
\caption{The SEE comparison of different schemes in in scenario C. } \label{fig14}
\end{figure}
\section{simulation  results}
In this section, numerical simulations are provided to evaluate the performance of our proposed scheme. Unless otherwise specified, the corresponding parameter values are summarized in Table I. As will be described later, we consider several network scenarios to study our proposed schemes.
\subsection{Single SUAV Case}
We first consider one UAV case, where   a single SUAV serves multiple legitimate users in the presence of one eavesdropper without the help of JUAV.  Fig.~\ref{fig2} shows the optimized UAV trajectories projected onto the horizontal plane for the different period $T$. As expected,   the UAV prefers moving closer to the legitimate users and far away from the eavesdropper  to  improve the system's secrecy rate. Meanwhile, we can also see that the UAV trajectories are smooth since it is in general less power-consuming\cite{zeng2018energy}.


To show the superiority of our proposed scheme  in terms of SEE of UAV systems, we consider the following benchmark schemes:

\begin{itemize}
\item \textbf{Optimized SEE maximization scheme:} This is our proposed scheme obtained from Algorithm 3 by jointly optimizing  the UAV transmit power, user scheduling and UAV trajectory.
\item \textbf{SEE maximization with circular path scheme:} For this scheme, the UAV flies with a circular path (initial trajectory). The SEE of UAV systems is obtained by jointly optimizing the UAV transmit power  and  user scheduling.
\item \textbf{UAV energy consumption minimization scheme:} For this scheme, we first obtain the optimized UAV trajectory via minimizing the UAV propulsion energy consumption. Then, with the  obtained  UAV trajectory, the secrecy rate is optimized by jointly optimizing the  UAV transmit power and  user scheduling.
\item \textbf{Secrecy rate maximization scheme:} The  UAV energy consumption is not optimized in this scheme. Our aim is to maximize the  secrecy rate  by jointly optimizing the UAV transmit power, user scheduling and UAV trajectory.
\end{itemize}
It is observed from Fig.~\ref{fig5} that  for our proposed scheme, the SEE  first monotonically increases, and then monotonically decreases  with period $T$. This is due to the fact that in the first phase, the incremental of UAV energy consumption is no larger than the incremental of secrecy rate  as period $T$ is small, thus  increases the system SEE. In the second phase, the incremental of UAV energy consumption is dramatically increased compared with  the incremental of secrecy rate as period $T$ becomes larger, thus decreases the system SEE. In addition, we can  see that our proposed scheme achieves significantly higher SEE as compared with the  benchmarks, which demonstrates the superiority of our proposed scheme.
\subsection{One SUAV and One JUAV Case}
In this subsection, we consider two UAVs case, where one SUAV and one JUAV simultaneously serve two legitimate users in the presence of one eavesdropper. Evidently,  the transmit power  of SUAV and JUAV should be carefully designed as the secrecy rate performance of UAV systems  will be degraded  by the  interference imposed  by JUAV.


We   plot the optimized SUAV trajectory  and JUAV  obtained from   Algorithm 3. It is  observed from Fig.~\ref{fig6} that  both optimized trajectories are rather similar to the circular trajectory, which has the  same result in \cite{zeng2018energy}. In addition,  the SUAV prefers moving closer to the legitimate user for data transmitting and JUAV prefers moving closer to eavesdropper to impose strong interference on eavesdropper. Fig.~\ref{fig7} shows that as the SUAV is closer to user 1, the SUAV will transmit information data to user 1 rather than user 2. Similarly, as the  SUAV moves closer to user 2, the SUAV  tends to transmit information data to user 2 rather than user 1. Fig.~\ref{fig8} shows  the power transmit  of SUAV and JUAV, as expected, when the SUAV moves from the legitimate user to eavesdropper, less power is allocated. Besides, more JUAV power is transmitted when the SUAV flies closer to the eavesdropper. This is because when the eavesdropper-SUAV channel  is good, the stronger jamming signal power  needs to be transmitted to interfere the eavesdropper and less power of SUAV is transmitted to prevent the data information  wiretapped by the  eavesdropper.

Fig.~\ref{fig9} depicts the secrecy rate performance with different period $T$. We compare the following two schemes, namely no cooperative JUAV scheme as well as  no  trajectory optimization scheme (circular trajectory). Clearly, the secrecy rate is monotonically increasing with time $T$. In addition, our proposed scheme achieves a significant higher secrecy rate than the other two benchmarks, which means the UAV trajectory has prominent impacts on the performance of secrecy rate. Also, for the no cooperative JUAV scheme, the system secrecy rate performance is poor, which indicates that JUAV indeed can bring the performance gain.

In the next, we have performed the new simulations  to show the UAV energy consumption  of our proposed SEE scheme as compared to other benchmarks. In Fig.~\ref{figtemp9}, it is observed that for  secrecy rate maximization scheme, the UAV consumes a large amount of propulsion energy compared with SEE scheme and UAV energy consumption minimization scheme. In addition, the gap between SEE scheme and UAV energy consumption minimization scheme is small, which indicates that the proposed SEE scheme  strikes an optimal balance between maximizing the secrecy achievable rate  and minimizing the UAV's propulsion energy consumption. In Fig.~\ref{fig10}, we compare the SEE achieved by the four schemes. Similar  results can be obtained from  Fig.~\ref{fig5}, the detailed explanations are  omitted here.
\subsection{Multi-SUAV and Multi-JUAV Case}
A more general case of two SUAVs and two JUAVs to serve multiple legitimate users in the presence of two  eavesdroppers is considered in this subsection. Fig.~\ref{fig11} depicts the optimized SUAV and JUAV trajectories obtained by using  Algorithm 3. It is  observed from Fig.~\ref{fig11} that the trajectories between SUAV 1 and JUAV 2 or (SUAV 2 and JUAV 1) tend to keep away to alleviate the co-channel interference. Meanwhile, the co-channel interference may be beneficial for the secure UAV system   by appropriately  imposing the jamming signal to the eavesdroppers. The corresponding SUAVs and JUAVs transmit power versus period $T$ are plotted in Fig.~\ref{fig12}. First, it is observed that when the
distance between legitimate user and  eavesdropper is not long (e.g., the legitimate user 1 and eavesdropper 1), the SUAV tends to move closer to legitimate user with less SUAV transmit power and meanwhile the JUAV transmits higher power. Second, when the distance between  legitimate user and  eavesdropper is  long (e.g., the legitimate user 2 and eavesdropper 1),  the SUAV tends to move closer to the legitimate user with higher   transmit power in order to improve the secrecy rate.

In Fig.~\ref{fig13}, the secrecy rate achieved by the various schemes  versus period $T$ is plotted. It is first found that our optimized trajectory scheme significantly outperforms the circular trajectory scheme as well as no cooperative JUAVs transmission scheme.  In Fig.~\ref{figtemp13}, We plot the UAV energy consumption with different schemes under different period time $T$. We can obtain   similar insights  in  Fig.~\ref{figtemp9}, which further demonstrate its correction. Furthermore, the system SEE performance against SEE maximization with circular path scheme, UAV energy consumption minimization scheme and secrecy rate maximization scheme are shown in Fig.~\ref{fig14}. It can be still observed that our  proposed scheme achieves significant gain compared with these three benchmarks in terms of SEE. In addition, we can  see that the trend of curves in Fig.~\ref{fig14} is similar with Fig.~\ref{fig10}, which indicates there exist a fundamental tradeoff between secrecy rate and UAV power consumption.
\section{Conclusion}
This paper investigates the energy-efficient multi-UAV enabled secure transmission wireless systems by considering the UAVs' propulsion energy consumption and users' secure rate simultaneously. We aim at maximizing the SEE of UAV systems by jointly optimizing the  user scheduling, transmit power and UAV trajectory. Then, an efficient three-layer iterative algorithm is proposed to solve the formulated non-convex and integer fractional problem  based on  block coordinate descent and Dinkelbach method, as well as SCA  techniques. Numerical results show that the UAV mobility is beneficial for achieving higher secure rate than the other benchmarks without considering trajectory optimization. Moreover, three useful insights are extracted from  numerical results.  First, our proposed  JUAVs-aided secrecy  rate maximization scheme  achieves significantly higher secrecy rate compared with no JUAVs-aided secure scheme. Second, the UAV-enabled  SEE does not   monotonically increase or decrease with period time $T$, in contrast, the UAV-enabled  SEE  is firstly increasing with period  $T$ and then decreasing with period time $T$. This is different with the secrecy rate maximization scheme and common throughput maximization as in \cite{wu2018Joint}. Third, our proposed SEE scheme gains significantly higher energy efficiency than that of the energy-minimization and secrecy rate maximization schemes.

There are still many other research directions can further extend this work. 1) We model  the A2G channel as free path loss for simplicity, the more practical channels such as Rician and Nakagami\text{-}m fading can be considered in the future work. 2) The design of UAV altitude can be further exploited, and how to efficiently optimize the joint UAV altitude, transmit power, user scheduling and UAV trajectory  is  also worthy of investigation. 3) This paper  considers the fixed wing UAV, the other types of UAVs such as the rotary wing UAV has the different energy consumption as well as UAV trajectory model, and it is still worthy of investigation. 4)  Some literatures  have paid attention to investigating the imperfect location of the eavesdropper scenario \cite{kang2019secrecy,zhou2018improving,zheng2014transmission}, and how to extend it in our  scenario is an interesting work.

\appendices
%

\section{Proof of Lemma  \ref{lemma2} }
For the sake simplicity, we first define a function $f\left( {{x_1}, \ldots ,{x_N}} \right) = {\log _2}\left( {\sum\limits_{i = 1}^N {{a_i}{x_i} + b} } \right)$, where ${x_i}$, ${a_i}$, and $b$ are all positive, $\forall i$. It is not difficult to verify that $f\left( {{x_1}, \ldots ,{x_N}} \right)$ is concave w.r.t. $x_i, \forall i$,  by checking its corresponding Hessian matrix. Recall that  the first-order Taylor expansion of a concave function is its global over-estimator \cite{boyd2004convex}, i.e., $f\left( {{x_1}, \ldots ,{x_N}} \right) \le f\left( {x_1^0, \ldots ,x_N^0} \right) + \sum\limits_{i = 1}^N {{{f'}_{{x_i}}}\left( {x_1^0, \ldots ,x_N^0} \right)\left( {{x_i} - x_i^0} \right)}$, where $x_i^0$, $\forall i$, denotes the  feasible point. Thus,  we have the  following inequality
\begin{align}
{\log _2}\left( {\sum\limits_{i = 1}^N {{a_i}{x_i} + b} } \right)& \le  {\log _2}\left( {\sum\limits_{i = 1}^N {{a_i}x_i^0 + b} } \right) \notag\\
&+ \sum\limits_{i = 1}^N {\frac{{{a_i}{\rm log}_2e}}{{\sum\limits_{j = 1}^N {{a_j}x_j^0 + b} }}} \left( {{x_i} - x_i^0} \right). \label{appendixconst1}
\end{align}

Therefore, \eqref{tildeup} follows from \eqref{appendixconst1} by setting $N={{{\cal M}_2}\backslash \left\{ {{m_2}} \right\} \cup {\cal M}_1}$, $a_i=h_{k_2,i}[n]$, $x_i=p_i[n]$, $x_i^0=p_i^r[n]$ and $b=\delta^2$. This completes the proof of Lemma \ref{lemma2}.

\ifCLASSOPTIONcaptionsoff
  \newpage
\fi

\bibliographystyle{IEEEtran}
\bibliography{TVT}

\end{document}